\documentclass[traditabstract]{aa}
\usepackage{url}
\usepackage{graphicx}
\usepackage{subfigure}
\usepackage{amsmath}
\usepackage{color}
\usepackage{listings} 
\usepackage{multirow}
\usepackage{float}
\usepackage{morefloats}
\usepackage{ifthen}
\usepackage{longtable}
\bibpunct{(}{)}{;}{a}{}{,} 
\usepackage[colorlinks=true,linkcolor=red,citecolor=blue, breaklinks=true]{hyperref}
\usepackage[switch, mathlines,]{lineno}
\usepackage[normalem]{ulem}
\usepackage{fancyvrb}

\usepackage{natbib,twoopt}
\makeatletter
\newcommandtwoopt{\citeads}[3][][]{\href{http://adsabs.harvard.edu/abs/#3}%
{\def\hyper@linkstart##1##2{}%
\let\hyper@linkend\@empty\citealp[#1][#2]{#3}}}
\newcommandtwoopt{\citepads}[3][][]{\href{http://adsabs.harvard.edu/abs/#3}%
{\def\hyper@linkstart##1##2{}%
\let\hyper@linkend\@empty\citep[#1][#2]{#3}}}
\newcommandtwoopt{\citetads}[3][][]{\href{http://adsabs.harvard.edu/abs/#3}%
{\def\hyper@linkstart##1##2{}%
\let\hyper@linkend\@empty\citet[#1][#2]{#3}}}
\newcommandtwoopt{\citeyearads}[3][][]%
{\href{http://adsabs.harvard.edu/abs/#3}
{\def\hyper@linkstart##1##2{}%
\let\hyper@linkend\@empty\citeyear[#1][#2]{#3}}}
\makeatother

\providecommand{\sorthelp}[1]{}
\newbox\tablebox    \newdimen\tablewidth
\def\leaderfil{\leaders\hbox to 5pt{\hss.\hss}\hfil}
\def\endPlancktablewide{\tablewidth=\textwidth 
    $$\hss\copy\tablebox\hss$$
    \vskip-\lastskip\vskip -2pt}
\def\tablenote#1 #2\par{\begingroup \parindent=0.8em
    \abovedisplayshortskip=0pt\belowdisplayshortskip=0pt
    \noindent
    $$\hss\vbox{\hsize\tablewidth \hangindent=\parindent \hangafter=1 \noindent
    \hbox to \parindent{$^#1$\hss}\strut#2\strut\par}\hss$$
    \endgroup}
\def\doubleline{\vskip 3pt\hrule \vskip 1.5pt \hrule \vskip 5pt}
\def\deg{\ifmmode^\circ\else$^\circ$\fi}
\def\pdeg{\ifmmode $\setbox0=\hbox{$^{\circ}$}\rlap{\hskip.11\wd0 .}$^{\circ}
          \else \setbox0=\hbox{$^{\circ}$}\rlap{\hskip.11\wd0 .}$^{\circ}$\fi}
\def\parcm{\rlap{.}$^{\scriptstyle\prime}$}

\def\Planck{\textit{Planck}}

\def\mo{\ifmmode^{-1}\else$^{-1}$\fi}
\def\,{\thinspace}

\def\WHzsr{\ifmmode $W\,Hz\mo\,sr\mo$\else W\,Hz\mo\,sr\mo\fi}
\def\mHz{\ifmmode $\,mHz$\else \,mHz\fi}
\def\GHz{\ifmmode $\,GHz$\else \,GHz\fi}
\def\mKs{\ifmmode $\,mK\,s$^{1/2}\else \,mK\,s$^{1/2}$\fi}
\def\muKs{\ifmmode \,\mu$K\,s$^{1/2}\else \,$\mu$K\,s$^{1/2}$\fi}
\def\muKRJs{\ifmmode \,\mu$K$_{\rm RJ}$\,s$^{1/2}\else \,$\mu$K$_{\rm RJ}$\,s$^{1/2}$\fi}
\def\muKHz{\ifmmode \,\mu$K\,Hz$^{-1/2}\else \,$\mu$K\,Hz$^{-1/2}$\fi}
\def\MJysr{\ifmmode \,$MJy\,sr\mo$\else \,MJy\,sr\mo\fi}
\def\KJysr{\ifmmode \,$kJy\,sr\mo$\else \,kJy\,sr\mo\fi}
\def\MJysrmK{\ifmmode \,$MJy\,sr\mo$\,mK$_{\rm CMB}\mo\else \,MJy\,sr\mo\,mK$_{\rm CMB}\mo$\fi}
\def\microns{\ifmmode \,\mu$m$\else \,$\mu$m\fi}

\def\muK{\ifmmode \,\mu$K$\else \,$\mu$\hbox{K}\fi}
\def\microK{\ifmmode \,\mu$K$\else \,$\mu$\hbox{K}\fi}
\def\muW{\ifmmode \,\mu$W$\else \,$\mu$\hbox{W}\fi}
\def\kms{\ifmmode $\,km\,s$^{-1}\else \,km\,s$^{-1}$\fi}
\def\cm2{\ifmmode $\,cm$^{-2}\else \,cm$^{-2}$\fi}
\def\kmsMpc{\ifmmode $\,\kms\,Mpc\mo$\else \,\kms\,Mpc\mo\fi}
\def\Mpc{\ifmmode $\,Mpc\mo$\else\,Mpc\mo\fi}
\DeclareMathAlphabet{\mathsc}{OT1}{cmr}{m}{sc}
\def\testbx{bx}%
\DeclareRobustCommand{\ion}[2]{%
\relax\ifmmode
\ifx\testbx\f@series
{\mathbf{#1\,\mathsc{#2}}}\else
{\mathrm{#1\,\mathsc{#2}}}\fi
\else\textup{#1\,{\mdseries\textsc{#2}}}%
\fi}

\newcommand{\planck}{{\it Planck\/}}

\newcommand{\healpix}{\ensuremath{\tt HEALPix}}
\newcommand{\xpol}{\ensuremath{\tt Xpol}}
\newcommand{\xpure}{\ensuremath{\tt Xpure}}
\newcommand{\polspice}{\ensuremath{\tt PolSpice}}

\newcommand{\kcmb}{\ensuremath{{\rm K}_{\rm CMB}}}
\newcommand{\mukcmb}{\ensuremath{{\mu \rm K}_{\rm CMB}}}
\newcommand{\mukcmbsq}{\ensuremath{{\mu \rm K}_{\rm CMB}}^2}

\newcommand{\NHI}{\ensuremath{N_\mathsc {Hi}}}
\newcommand{\NHIi}{\ensuremath{N_\mathsc {Hi}^i}}
\newcommand{\hi}{\ensuremath{\mathsc {Hi}}}

\newcommand{\Nside}{\ensuremath{N_{\rm side}}}

\newcommand{\E}{{\ifmmode E\else $E$\fi}}
\newcommand{\B}{{\ifmmode B\else $B$\fi}}   
\newcommand{\EE}{{\ifmmode EE\else $EE$\fi}}    
\newcommand{\BB}{{\ifmmode BB\else $BB$\fi}}    
\newcommand{\TE}{{\ifmmode TE\else $TE$\fi}}    
\newcommand{\TB}{{\ifmmode TB\else $TB$\fi}}    
\newcommand{\Bt}{\ensuremath{\vec{B}_{\rm turb}}}

\newcommand{\Bo}{\ensuremath{\vec{B}_{\rm ord}}}
\newcommand{\BT}{\ensuremath{\vec{B}}}
\newcommand{\Bpos}{\ensuremath{\vec{B}_{\rm POS}}}

\newcommand{\uBo}{\ensuremath{\hat{\vec{B}}_{\rm ord}}}

\newcommand{\uBt}{\ensuremath{\hat{\vec{B}}_{\rm turb}}}
\newcommand{\Dmodel}{\ensuremath{D_{353}}}

\newcommand{\dlxx}{\ensuremath{{\cal D}_{\ell}^{XX}}}
\newcommand{\dltt}{\ensuremath{{\cal D}_{\ell}^{TT}}}
\newcommand{\dlee}{\ensuremath{{\cal D}_{\ell}^{EE}}}
\newcommand{\dlbb}{\ensuremath{{\cal D}_{\ell}^{BB}}}
\newcommand{\dlte}{\ensuremath{{\cal D}_{\ell}^{TE}}}

\newcommand{\mlxx}{\ensuremath{{\cal M}_{\ell}^{XX}}}
\newcommand{\mltt}{\ensuremath{{\cal M}_{\ell}^{TT}}}
\newcommand{\mlee}{\ensuremath{{\cal M}_{\ell}^{EE}}}
\newcommand{\mlbb}{\ensuremath{{\cal M}_{\ell}^{BB}}}
\newcommand{\mlte}{\ensuremath{{\cal M}_{\ell}^{TE}}}
\newcommand{\mltb}{\ensuremath{{\cal M}_{\ell}^{TB}}}
\newcommand{\mleb}{\ensuremath{{\cal M}_{\ell}^{EB}}}

\newcommand{\resdiff}{\Delta}

\begin{document}

\title{Modelling and simulation of large-scale polarized dust emission over the southern Galactic cap using the GASS \hi\ data}
\author{T. Ghosh\thanks{Corresponding author:  tghosh@caltech.edu} \inst{1,2},  F. Boulanger \inst{2}, P. G. Martin \inst{3},  A. Bracco \inst{4,2}, F. Vansyngel \inst{2},  J. Aumont \inst{2}, J. J. Bock\inst{1,5},   \\  O. Dor\'{e} \inst{1,5}, U. Haud \inst{6}, P. M. W. Kalberla \inst{7}, and P. Serra\inst{1,5}}

\institute{California Institute of Technology, Pasadena, California, U.S.A.\\
\and
Institut d'Astrophysique Spatiale, CNRS (UMR8617) Universit\'{e} Paris-Sud 11, B\^{a}timent 121, Orsay, France\\
\and
CITA, University of Toronto, 60 St. George St., Toronto, ON M5S 3H8, Canada\\
\and
Laboratoire AIM, IRFU/Service d'Astrophysique - CEA/DSM - CNRS - Universit\'{e} Paris Diderot, B\^{a}t. 709, CEA-Saclay, F-91191 Gif-sur-Yvette Cedex, France\\
\and
Jet Propulsion Laboratory, California Institute of Technology, 4800 Oak Grove Drive, Pasadena, California, U.S.A.\\
\and
Tartu Observatory, 61602 T\~{o}ravere, Tartumaa, Estonia\\
\and
Argelander-Institut f\"{u}r Astronomie, Universit\"{a}t Bonn, Auf dem H\"{u}gel 71, D-53121 Bonn, Germany\\
            }

\date{Received ....; accepted ....}

\abstract{ The \planck\ survey has quantified polarized Galactic foregrounds and established that they are a main limiting factor in the quest for the cosmic microwave background (CMB) $B$-mode signal induced by primordial gravitational waves during cosmic inflation. Accurate separation of the Galactic foregrounds therefore binds this quest to our understanding of the magnetized interstellar medium (ISM). The two most relevant empirical results from analysis of \planck\ data are line of sight depolarization arising from fluctuations of the Galactic magnetic field orientation and alignment of filamentary dust structures with the magnetic field at high Galactic latitude. Furthermore, \planck\ and \hi\ emission data in combination indicate that most of the filamentary dust structures are in the cold neutral medium. The goal of this paper is to test whether these salient observational results, taken together, can account fully for the statistical properties of the dust polarization over a selected low column density region comprising 34\,\% of the southern Galactic cap ($b \le -30\deg$). To do this, we construct a dust model that incorporates \hi\ column density maps as tracers of the dust intensity structures and a phenomenological description of the Galactic magnetic field. By adjusting the parameters of the dust model, we were able to reproduce the \planck\ dust observations at 353\,GHz in the selected region. Realistic simulations of the polarized dust emission enabled by such a dust model are useful for testing the accuracy of component separation methods, studying non-Gaussianity, and constraining the amount of decorrelation with frequency. }

\titlerunning{Modelling and simulation of polarized dust emission} 
\authorrunning{T. Ghosh et al.} 
\maketitle


\section{Introduction} 

An intense focus in current observational cosmology is detection of the CMB $B$-mode signal induced by primordial gravitational waves during the inflation era \citep{Starobinsky:1979,Fabbri:1983,Abbott:1984}. The BICEP2 experiment\footnote{\url{http://bicepkeck.org}} puts an upper bound on the amplitude of the CMB $B$-mode signal, parameterized by a tensor-to-scalar ratio ($r$) at a level of $r\,<\,0.09$ (95\,\% confidence level \citep{BK:2016}. Therefore, the major challenge for a BICEP2-like experiment with limited frequency coverage is to detect such an incredibly faint CMB $B$-mode signal in the presence of foreground Galactic contamination \citep{Betoule:2009}. The dominant polarized foreground above 100\,GHz come from thermal emission by aligned aspherical dust grains \citep{planck2014-XXII, planck2014-a12}. Unlike the CMB, the polarized dust emission is distributed non-uniformly on the sky with varying column density over which are summed contributions from multiple components with different dust composition, size, and shape and with different magnetic field orientation (see reviews by \citealt{Prunet:1999,Lazarian:2008}). Acknowledging these complexities, the goal of this paper is to model the principal effects, namely multiple components with different magnetic field orientations, to produce realistic simulated maps of the polarized dust emission. 

Our knowledge and understanding of the dust polarization has improved significantly in the submillimetre and microwave range through exploitation of data from the \planck\ satellite\footnote{\planck\ (\url{http://www.esa.int/Planck}) is a project of the European Space Agency (ESA) with instruments provided by two scientific consortia funded by ESA member states and led by Principal Investigators from France and Italy, telescope reflectors provided through a collaboration between ESA and a scientific consortium led and funded by Denmark, and additional contributions from NASA (USA).} \citep{planck2014-a01,planck2014-a12,planck2014-XIX}, which provides important empirical constraints on the models. The \planck\ satellite has mapped the polarized sky at seven frequencies between 30 and 353\,GHz \citep{planck2014-a01}. The \planck\ maps are available in \healpix\footnote{\url{http://healpix.sourceforge.net}} format \citep{Gorski:2005} with resolution parameter labelled with the \Nside\ value. The general statistical properties of the dust polarization in terms of angular power spectrum at intermediate and high Galactic latitudes from 100 to 353\,GHz are quantified in \citet{planck2014-XXX}.  

Two unexpected results from the \planck\ observations were an asymmetry in the amplitudes of the angular power spectra of the dust $E$- and $B$- modes ($EE$ and $BB$ spectra, respectively) and a positive temperature E-mode ($TE$) correlation at 353\,GHz.  Explanation of these important features was explored by \citet{planck2015-XXXVIII} through a statistical study of the filamentary structures in the \Planck\ Stokes maps and \citet{Caldwell:2016} discussed them in the context of magnetohydrodynamic interstellar turbulence.  They are a central focus of the modelling in this paper.

The  \planck\ 353\,GHz polarization maps \citep{planck2014-XIX} have the best signal-to-noise ratio and enable study of the link between the structure of the Galactic magnetic field (GMF) and dust polarization properties. A large scatter of the dust polarization fraction ($p$) for total column density $N_{\rm H} < 10^{22}\, \text{cm}^{-2}$ at low and intermediate Galactic latitudes is reported in \citet{planck2014-XIX}. Based on numerical simulations of anisotropic magnetohydrodynamic turbulence in the diffuse interstellar medium (ISM), it has been concluded that the large scatter of $p$ in the range $2\times10^{21}\, \text{cm}^{-2} < N_{\rm H} < 2\times10^{22}\, \text{cm}^{-2}$ comes mostly from fluctuations in the GMF orientation along the line of sight (LOS) rather than from changes in grain shape or the efficiency of the grain alignment \citep{planck2014-XX}. To account for the observed scatter of $p$ in the high latitude southern Galactic sky ($b \le - 60 \deg$), a finite number of polarization layers was introduced \citep{planck2016-XLIV}. The orientation of large-scale GMF with respect to the plane of the sky (POS) also plays a crucial role in explaining the scatter of $p$ \citep{planck2014-XX}. 
  
The  \planck\ polarization data reveal a tight relationship between the orientation of the dust intensity structures and the magnetic field projected on the plane of the sky (\Bpos, \citealt{planck2014-XXXII, planck2015-XXXV, planck2015-XXXVIII}). In particular, the dust intensity structures are preferentially aligned with \Bpos\ in the diffuse ISM \citep{planck2014-XXXII}.  Similar alignment is also reported between the Galactic Arecibo L-Band Feed Array (GALFA, \citealt{Peek:2011}) \hi\ filaments and \Bpos\ derived either from starlight polarization \citep{Clark:2014} or \planck\ polarization data \citep{Clark:2015}. Furthermore, a detailed all-sky study of the \hi\ filaments, combining the Galactic All Sky Survey (GASS, \citealt{McClure:2009}) and the Effelsberg Bonn \hi\ Survey (EBHIS, \citealt{Winkel:2016}), shows that most of the filaments occur in the cold neutral medium (CNM, \citealt{Kalberla:2016}). The \hi\ emission in the CNM phase has a line profile with $1\sigma$ velocity dispersion $3 \kms$ (FWHM $7 \kms$) in the solar neighbourhood \citep{Heiles:2003} and filamentary structure on the sky, which might result from projection effects of gas organized primarily in sheets \citep{Heiles:2005,Kalberla:2016}. Alignment of CNM structures with \Bpos\ from \planck\ has been reported towards the north ecliptic pole, among the targeted fields of the Green Bank Telescope \hi\ Intermediate Galactic Latitude Survey (GHIGLS, \citealt{Martin:2015}). Such alignment can be induced by shear strain of gas turbulent velocities stretching matter and the magnetic field in the same direction  \citep{Hennebelle:2013,Inoue:2016}.

Alignment between the CNM structures and \Bpos\ was not included in the pre-\planck\ dust models, for example Planck Sky Model (PSM, \citealt{PSM:2013}) and FGPol \citep{ODea:2012}. The most recent version of post-\planck\ PSM uses a filtered version of the \planck\ 353\,GHz polarization data as the dust templates \citep{planck2014-a14}. Therefore, in low signal-to-noise regions the PSM dust templates are not reliable representations of the true dust polarized sky. In \citet{planck2015-XXXVIII} it was shown that preferential alignment of the dust filaments with \Bpos\ could account for both the $E$-$B$ asymmetry in the amplitudes of the dust \dlee\ and \dlbb\  angular power spectra\footnote{${\cal D}^{EE,BB}_{\ell} \equiv \ell (\ell+1) C^{EE,BB}_{\ell}/(2\pi$).} over the multipole range $40 < \ell < 600$ and the amplitude ratio $\dlte/\dlee$, as measured by \citet{planck2014-XXX}. \citet{Clark:2015} showed that the alignment of the \Bpos\ with the \hi\ structures can explain the observed $E$-$B$ asymmetry.

Recently, \citet{Vansyngel:2016} have extended the approach introduced in \citet{planck2016-XLIV} using a parametric model to account for the observed dust power spectra at intermediate and high Galactic latitudes. Our work in this paper is complementary, incorporating additional astrophysical data and insight and focusing on the cleanest sky area for CMB $B$-mode studies from the southern hemisphere.  

In more detail, the goal of this paper is to test whether preferential alignment of the CNM structures with \Bpos\ together with fluctuations in the GMF orientation along the LOS can account fully for a number of observed statistical properties of the dust polarization over the southern low column density sky. These statistical properties include the scatter of the polarization fraction $p$, the dispersion of the polarization angle $\psi$ around the large-scale GMF, and the observed dust power spectra \dlee, \dlbb, and \dlte\ at 353\,GHz.  In doing so we extend the analysis of  \citet{planck2015-XXXVIII} and \citet{planck2016-XLIV} to sky areas in which the filaments have very little contrast with respect to the diffuse background emission. Within the southern Galactic cap (defined here to be $b \le -30\deg$), in which dust and gas are well correlated \citep{Boulanger:1996,planck2013-p06b} and the GASS \hi\ emission is a good tracer of the dust intensity structures \citep{planck2013-XVII}, we focus our study on the low column density portion. We choose to model the dust polarization sky at 1\deg\ resolution, which corresponds to retaining multipoles $\ell \le 160$ where the Galactic dust dominates over the CMB $B$-mode signal. Also, most of the ground-based CMB experiments are focusing at a degree scale for the detection of the recombination bump of the CMB $B$-mode signal.
Our dust polarization model should be a useful product for the CMB community.  It covers 34\,\% of the southern Galactic cap, a 3500 deg$^2$ region containing the cleanest sky accessible to ongoing ground and balloon-based CMB experiments: ABS \citep{Essinger-Hileman:2010},  Advanced ACTPol \citep{Henderson:2016},  BICEP/Keck \citep{Ogburn:2012}, CLASS \citep{Essinger-Hileman:2014}, EBEX \citep{Reichborn:2010}, Polarbear \citep{Kermish:2012}, Simons array \citep{Suzuki:2016}, SPIDER \citep{Crill:2008}, and SPT3G \citep{Benson:2014}. 

This paper is organized as follows. Section~\ref{sec:data} introduces the \planck\ observations and GASS \hi\ data used in our analysis, along with a selected sky region within the southern Galactic cap, called SGC34. In Sect.~\ref{sec:obser}, we summarize the observational dust polarization properties in SGC34 as seen by \planck.  A model to simulate dust polarization using the GASS \hi\ data and a phenomenological description of the GMF is detailed in Sect.~\ref{sec:methodology}. The procedure to simulate the polarized dust sky towards the southern Galactic cap and the choice of the dust model parameters are discussed in Sect.~\ref{sec:model_params}. Section~\ref{sec:results} presents the main results where we statistically compare the dust model with the observed \planck\ data. In Sect.~\ref{sec:prediction}, we make some predictions using our dust model. We end with an astrophysical perspective in Sect.~\ref{sec:interpretation} and a short discussion and summary of our results in Sect.~\ref{sec:concl}.

\section{Data sets used and region selection}
\label{sec:data}

\subsection{\planck\ 353\,GHz data}
\label{sec:2.1}

In this paper we used the publicly available\footnote{\url{http://www.cosmos.esa.int/web/planck/pla}} maps from the \planck\ 2015 polarization data release (PR2), specifically at 353\,GHz \citep{planck2014-a01}.\footnote{These maps are expressed in thermodynamic units (\kcmb).  For this passband, multiply by 287.35 to convert to \MJysr.} The data processing and calibration of the \planck\ High Frequency Instrument (HFI) maps are described in \citet{planck2014-a09}. We used multiple \planck\ 353\,GHz polarization data sets: the two yearly surveys, Year 1 (Y1) and Year 2 (Y2), the two half-mission maps, HM1 and HM2,  and the two detector-set maps, DS1 and DS2, to exploit the statistical independence of the noise between them and to test for any systematic effects.\footnote{ The noise is partially correlated between the two half-ring maps  \citep{planck2013-p03} and so those maps are not used in our analysis.} The 353\,GHz polarization maps are corrected for zodiacal light emission at the map-making level \citep{planck2014-a09}. These data sets are described in full detail in \citet{planck2014-a09}. 

All of the 353\,GHz data sets used in this study have an angular resolution of FWHM 4\parcm82 and are projected on a \healpix\ grid with a resolution parameter $\Nside=2048$ (1\parcm875 pixels). We corrected for the main systematic effect, that is leakage from intensity to polarization, using the global template fit, as described in \citet{planck2014-a09}, which accounts for the monopole, calibration, and bandpass mismatches. To increase the signal-to-noise ratio of the \planck\ polarization measurements over SGC34, we smoothed the \planck\ polarization maps to 1\deg\ resolution using a Gaussian approximation to the \planck\ beam and reprojected to a \healpix\ grid with \Nside=128 (30\arcmin\ pixels).  We did not attempt to correct for the CMB polarization anisotropies at 1\deg\ resolution because they are negligibly small compared to the dust polarization at 353\,GHz.

The dust intensity at 353\,GHz, \Dmodel, is calculated from the latest publicly available \planck\ thermal dust emission model \citep{planck2016-XLVIII}, which is a modified blackbody (MBB) fit to maps at $\nu \ge$ 353\,GHz from which the cosmic infrared background anisotropies (CIBA) have been removed using the generalized linear internal combination (GNILC) method and zero-level offsets (e.g. the CIB monopole) have been removed by correlation with \hi\ maps at high Galactic latitude. By construction, this dust intensity map is corrected for zodiacal light emission, CMB anisotropies, and resolved point sources as well as the CIBA and the CIB monopole.  As a result of the GNILC method and the MBB fitting, the  noise in the \Dmodel\ map is lower than in the Stokes $I$ frequency map at 353\,GHz. We also smoothed the \Dmodel\ map to the common resolution of 1\deg, taking into account the effective beam resolution of the map.

\begin{figure*}
\includegraphics[width=19cm]{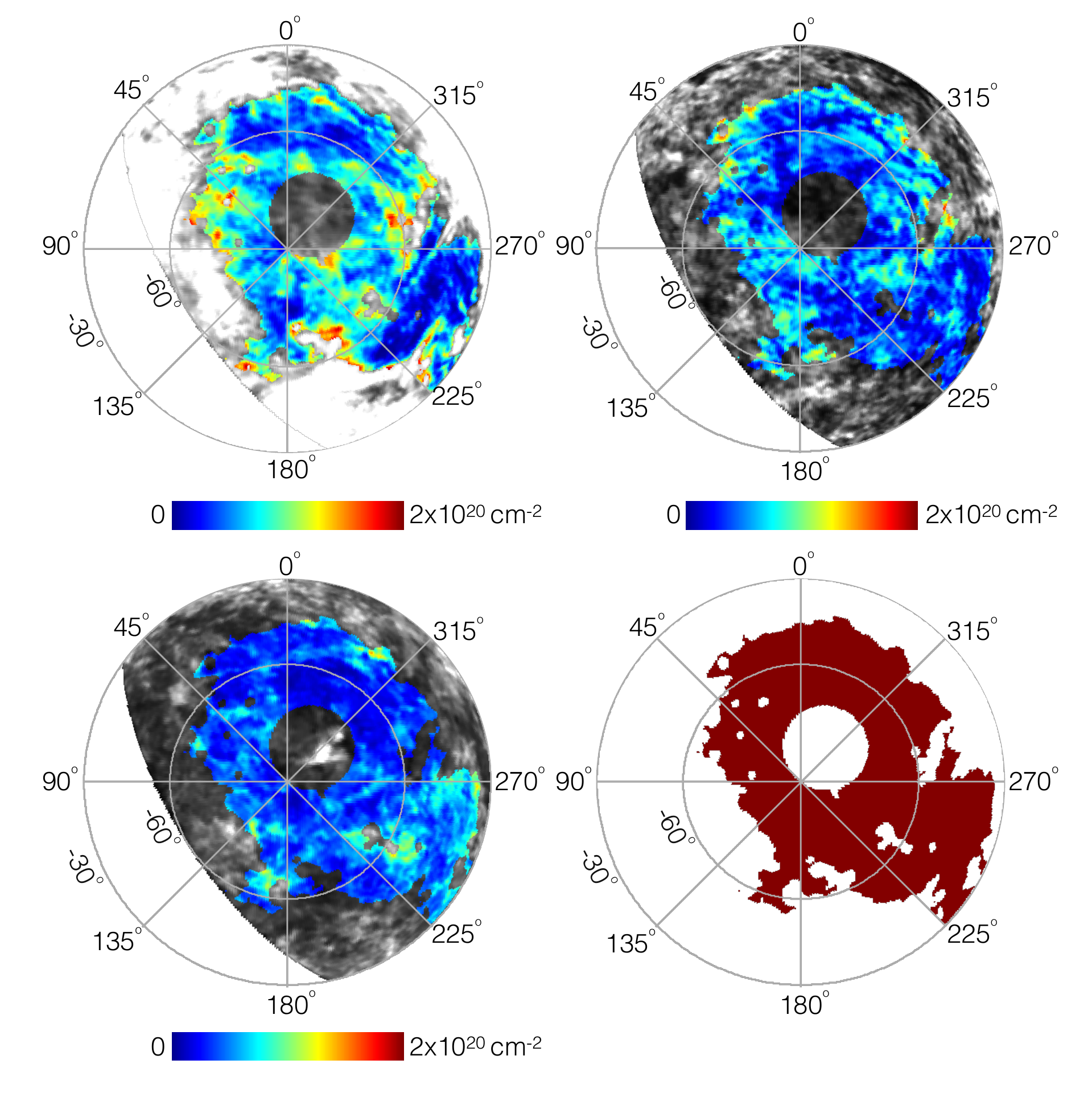} 
\caption{\NHI\ maps of the three Gaussian-based \hi\ phases: CNM (\textit{top left}), UNM (\textit{top right}), and WNM (\textit{bottom left}) over the portion of the southern Galactic cap ($b \le -30\deg$) covered by the GASS \hi\ survey, at 1\deg\ resolution.  Coloured region corresponds to the selected region SGC34 (\textit{bottom right}). Orthographic projection in Galactic coordinates: latitudes and longitudes marked by lines and circles respectively.}
\label{fig:4.2.1}
\end{figure*}

\subsection{GASS \hi\ data}
\label{sec:2.3}

We use the GASS \hi\ survey\footnote{\href{https://www.astro.uni-bonn.de/hisurvey/gass}{https://www.astro.uni-bonn.de/hisurvey/gass}} \citep{McClure:2009,Kalberla:2010,Kalberla:2015} which mapped the southern sky (declination $\delta < 1\deg$) with the Parkes telescope at an angular resolution of FWHM 14\parcm4.  The rms uncertainties of the brightness temperatures per channel are at a level of $57$\,mK at $\delta v= 1 \kms$.  The GASS survey is the most sensitive and highest angular resolution survey of \hi\ emission over the southern sky.  The third data release version has improved performance by the removal of residual instrumental problems along with stray radiation and radio-frequency interference, which can be major sources of error in \hi\ column density maps \citep{Kalberla:2015}. The GASS \hi\ data are projected on a \healpix\ grid with $\Nside = 1024$ (3\parcm75 pixels).

For this optically thin gas, the brightness temperature spectra $T_b$ of the GASS \hi\ data are decomposed into multiple Gaussian components  \citep{Haud:2007,Haud:2013} as
\begin{equation}
T_b= \sum_i T_0^i \, \exp\left[-\frac{1}{2} \left(\frac{v_{\rm LSR}-v_{\rm c}^i}{\sigma^i} \right)^2 \right]  \ ,
\end{equation}
where the sum is over all Gaussian components for a LOS, $v_{\rm LSR}$ is the velocity of the gas relative to the local standard of rest (LSR), and $T_0^i$, $v_{\rm c}^i$, and $\sigma^i$ are the peak brightness temperature (in K), central velocity, and $1\sigma$ velocity dispersion of the $i^{th}$ Gaussian component, respectively. Although the Gaussian decomposition does not provide a unique solution, it is correct empirically. \citet{Haud:2013} used the GASS survey at \Nside=1024 and decomposed the observed 6\,655\,155 \hi\ profiles into 60\,349\,584 Gaussians. In this study, we use an updated (but unpublished) Gaussian decomposition of the latest version of the GASS \hi\ survey (third data release). The intermediate and high velocity gas over the southern Galactic cap has negligible dust emission (see discussion in \citealt{planck2013-XVII}) and in any case lines of sight with uncorrelated dust and \hi\ emissions are removed in our masking process (Sect.~\ref{regionselection}).

Broadly speaking, the Gaussian \hi\ components represent the three conventional phases of the ISM.  In order of increasing velocity dispersion these are: (1) a cold dense phase, the cold neutral medium (CNM), with a temperature in the range $\sim\, 40-200$\,K; (2) a thermally unstable neutral medium (UNM), with a temperature in the range $\sim \, 200-5000$\,K; and (3) a warm diffuse phase, the warm neutral medium (WNM), with a temperature in the range $\sim\,5000-8000$\,K. Based on sensitive high-resolution Arecibo observations, \citet{Heiles:2003} found that a significant fraction of the \hi\ emission is in the UNM phase at high Galactic latitude. Numerical simulations of the ISM suggest that the magnetic field and dynamical processes like turbulence could drive \hi\ from the stable WNM and CNM phases into the UNM phase \citep{Audit:2005, Hennebelle:2014, Saury:2014}.

The total \hi\ column density along the LOS, \NHI, in units of $10^{18}\,\cm2$, is obtained by integrating the brightness temperature over velocity.  We can separate the total \NHI\ into the three different phases of the ISM based on the velocity dispersion of the fitted Gaussian. Thus
\begin{align}
\NHI & =  1.82\times \int  T_b \, dv_{\rm LSR}  \nonumber \\
&= 1.82 \times \int  \sum_i T_0^i \, \exp\left[-\frac{1}{2} \left(\frac{v_{\rm LSR}-v_{\rm c}^i}{\sigma^i} \right)^2 \right] dv_{\rm LSR}  \nonumber  \\
& =  1.82 \times \sqrt{2\pi}  \sum \left ( T_0^i \, \sigma^i \, w_{\rm c} +  T_0^i \, \sigma^i \, w_{\rm u}  + T_0^i \, \sigma^i  \, w_{\rm w} \right)  \nonumber \\
& \equiv \NHI^{\rm c} + \NHI^{\rm u} + \NHI^{\rm w} \ ,
\label{eq:3.1.1}
\end{align}
where $w_{\rm c}$, $w_{\rm u}$, and $w_{\rm w}$ are the weighting factors to select the CNM, UNM, and WNM components, respectively, and $\NHI^{\rm c}$, $\NHI^{\rm u}$, and $\NHI^{\rm w}$ are the CNM, UNM, and WNM column densities, respectively. The weighting factors are defined as
\begin{align}
w_{\rm c}&=\left\{\begin{array}{ll}
1 & \sigma < \sigma_{\rm c} \\
\frac{1}{2}\left[1 + \cos\left( \frac{\pi}{g} \frac{\sigma -\sigma_{\rm c}}{\sigma_{\rm c}}\right)\right] & \sigma_{\rm c} \le \sigma \le \sigma_{\rm c} (1+g)\\
0 & \sigma > \sigma_{\rm c} (1+g)
\end{array}\right.  \nonumber \\
w_{\rm w}&=\left\{ \begin{array}{ll}
0 & \sigma < \sigma_{\rm u} \\
\frac{1}{2}\left[1 - \cos\left( \frac{\pi}{g} \frac{\sigma -\sigma_{\rm u}}{ \sigma_{\rm u}}\right)\right] & \sigma_{\rm u} \le \sigma < \sigma_{\rm u} (1+g)\\
1 & \sigma \ge \sigma_{\rm u} (1+g) \\
\end{array}\right. \\
w_{\rm u} &= 1 - w_{\rm c} - w_{\rm w} \ . \nonumber
\end{align}
We use overlapping weighting factors to avoid abrupt changes in the assignment of the emission from a given \hi\ phase to another, adopting $g = 0.2$.  There is a partial correlation between two neighbouring \hi\ components.  To select local velocity gas, we included only Gaussian components with $|v_{\rm c}^i| \, \le \, 50\, \kms$.

For our analysis we also smoothed these \NHI\ maps to the common resolution of 1\deg, taking into account the effective beam resolution of the map.  Figure~\ref{fig:4.2.1} shows the separation of the total \NHI\ into the three Gaussian-based \hi\ phases, CNM, UNM, and WNM.  The weights were calculated with $\sigma_{\rm c}=7.5\, \kms$ and $\sigma_{\rm u}=10\, \kms$ as adopted in Sect.~\ref{choiceparams}.   The total \NHI\ map obtained by summing these component phases is very close to the LVC map in Fig.~1 of \citet{planck2013-XVII} obtained by integrating over the velocity range defined by Galactic rotation, with dispersion only $0.012 \times10^{20}\cm2$ in the difference or 0.8\,\% in the fractional difference over the selected region SGC34 defined below.

To build our CNM map we use an upper limit to the velocity dispersion  ($\sigma_c\,=\,7.5\kms$) larger than the conventional line width of cold \hi\ gas \citep{Wolfire:2003, Heiles:2005, Kalberla:2016}. The line width can be characterised by a Doppler temperature $T_{\rm D}$. Because the line width includes a contribution from turbulence, $T_{\rm D}$ is an upper limit to the gas kinetic temperature. In \citet{Kalberla:2016}, the median Doppler temperature of CNM gas is about $T_{\rm D}=220$\,K, corresponding to a velocity dispersion of about $\sigma_c=1.3\kms$, while the largest value of $T_{\rm D}=1100$\,K corresponds to $\sigma_c=3.0\kms$. The value of  $\sigma_c=7.5\kms$ in our model follows from the model fit to the  \planck\ polarization data (the observed $E-B$ asymmetry and the $\dlte/\dlee$ ratio over SGC34, see Sect.~\ref{choiceparams}).  Thus our CNM component ($\NHI^{\rm c}$) includes some UNM gas. To investigate the impact of this on our modelling, we divided our CNM component map into two parts corresponding to components with velocity dispersions above and below $\sigma_c=3.0\,\kms$. We then repeated our data fitting with these two separate maps replacing our single CNM component map and found no significant difference in the model results.


\subsection{Selection of the region SGC34}
\label{regionselection}

For our analysis we select a low dust column density region of the sky within the southern Galactic cap.  As a first cut, we retain only the sky pixels that have \NHI\ below a threshold value, $\NHI\,\le \, 2.7\times10^{20}\,\cm2$. Molecular gas is known from UV observations to be negligible at this threshold \NHI\ value \citep{Gillmon:2006}. We adopt a second mask, from \citet{planck2013-XVII}, to avoid sky pixels that fall off the main trend of gas-dust correlation between the GASS \hi\ column density and \planck\ HFI intensity maps, in particular applying a threshold of 0.21\MJysr\ ($3\sigma$ cut, see Figure 4 of \citealt{planck2013-XVII}) on the absolute value of the residual emission at 857\,GHz after subtraction of the dust emission associated with \hi\ gas. We smooth the second mask to Gaussian 1\deg\ FWHM resolution and downgrade to \Nside=128 \healpix\ resolution. Within this sky area, we also mask out a $10\deg$ radius patch centred around $(l,b) = (324\pdeg05, -79\pdeg55)$ to avoid \hi\ emission from the Magellanic stream, whose mean radial velocity for this specific sky patch is within the Galactic range of velocities \citep{Nidever:2010}. 

After masking, we are left with a 3500 deg$^2$ region comprising 34\,\% of the southern Galactic cap, referred to hereafter as SGC34. This is presented in Fig.~\ref{fig:4.2.1}. The mean \NHI\ over SGC34 is $\langle \NHI \rangle = 1.58\times10^{20}\cm2$. For the adopted values of $\sigma_{\rm c}$ and $\sigma_{\rm u}$, the mean column densities of the three ISM phases (Fig.~\ref{fig:4.2.1}) are $\langle \NHI^{\rm c}\rangle=0.66\times10^{20}\cm2$, $\langle \NHI^{\rm u}\rangle=0.49\times10^{20}\cm2$, and $\langle \NHI^{\rm w}\rangle=0.43\times10^{20}\cm2$. SGC34 has a mean dust intensity of $\langle \Dmodel \rangle = 54\,\KJysr$.  Correlating the total \NHI\  map with the \Dmodel\ map, we find an emissivity (slope) $32\, \KJysr  (10^{20}\,\cm2)^{-1}$ at 353\,GHz (Sect.~\ref{choiceparams}) and an offset of just 3\,\KJysr, which we use below as a $1\sigma$ uncertainty on the zero-level of the Galactic dust emission.

For computations of the angular power spectrum, we apodized the selection mask by convolving it with a 2\deg\ FWHM Gaussian. The effective sky coverage of SGC34 as defined by the mean sky coverage of the apodized mask is $f_{\rm sky}^{\rm eff}$\,=\,0.085 (8.5\,\%), meaning that the smoothing extends the sky region slightly but conserves the effective sky area.

\subsection{Power spectrum estimator}

We use the publicly available \xpol\ code \citep{Tristram:2005} to compute the binned angular power spectra over a given mask. A top-hat binning in intervals of $20\,\ell$ over the range $40 < \ell < 160$ is applied throughout this analysis. The upper cutoff $\ell_{\rm max}=160$ is set by 1\deg\ smoothed maps used throughout this paper. The lower cutoff $\ell_{\rm min}=40$ is chosen to avoid low $\ell$ systematic effects present in the publicly available \planck\ polarization data \citep{planck2014-a10}. The \xpol\ code corrects for the incomplete sky coverage, leakage from $E$- to $B$-mode polarization, beam smoothing, and pixel window function. Other codes such as \polspice\ \citep{Chon:2004} and \xpure\ \citep{Grain:2009} also compute angular power spectra from incomplete sky coverage. We prefer \xpol\ over other codes because it directly computes the analytical error bars of the power spectrum without involving any Monte-Carlo simulations. The final results of this paper do not depend on the \xpol\ code, as the same mean power spectra are returned by other power spectrum estimators.

\section{\planck\ dust polarization observations over the selected region SGC34}
\label{sec:obser}

In this section we derive the statistical properties of the \planck\ dust polarization over SGC34. These statistical properties include normalized histograms from pixel space data, specifically of $p^2$ and of $\psi$ with respect to the local polarization angle of the large-scale GMF, and the polarized dust power spectra in the harmonic domain.

\begin{table*}[tb]
\begingroup
\newdimen\tblskip \tblskip=5pt
\caption{Dust polarization fraction $p$ statistics at 353\,GHz and the best-fit mean direction of the large-scale GMF over SGC34 using multiple subsets of the \planck\ data. }
\label{tab:1} 
\nointerlineskip
\vskip -3mm
\footnotesize
\setbox\tablebox=\vbox{
   \newdimen\digitwidth 
   \setbox0=\hbox{\rm 0} 
   \digitwidth=\wd0 
   \catcode`*=\active 
   \def*{\kern\digitwidth}
   \newdimen\signwidth 
   \setbox0=\hbox{+} 
   \signwidth=\wd0 
   \catcode`!=\active 
   \def!{\kern\signwidth}
\halign{
\hbox to 2.5cm{#\hfil}\tabskip 2.0em&
\hfil #\hfil &
\hfil #\hfil &
\hfil #\hfil &
\hfil #\hfil &
\hfil #\hfil &
\hfil #\hfil &
\hfil #\hfil\tabskip=0pt\cr
\noalign{\doubleline}
 \omit Dataset \hfil & \multispan{3}\hfil polarization fraction ($p_{\rm d}$) \hfil & $f (p_{\rm d} > 25\,\%$)   & subsets & $l_0$ & $b_0$ \cr
  \omit  \hfil & median & mean & maximum &  &  & &  \cr
\noalign{\doubleline}
Detsets &  8.0\,\% & 11.0\% & 33.4\,\% & 1.2\,\% & DS1 & 77\pdeg3 &  25\pdeg5 \cr
              &        &       &         &       & DS2 &70\pdeg4 & 21\pdeg8 \cr
\noalign{\vskip 4pt\hrule\vskip 6pt}
HalfMissions & 8.5\,\%  & 11.6\,\% & 35.4\,\% &  1.9\,\% & HM1 & 73\pdeg2 &  23\pdeg0\cr
              &        &       &         &       & HM2 & 73\pdeg8& 24\pdeg0 \cr
\noalign{\vskip 4pt\hrule\vskip 6pt}
Years & 8.3\,\% &11.3\,\% & 34.6\,\% & 1.4\,\% & YR1 &73\pdeg6 & 23\pdeg2 \cr
              &        &       &         &       & YR2 & 73\pdeg0 & 23\pdeg3 \cr
\noalign{\doubleline}
average & 8.3\,\% & 11.3\,\% & -- & -- & -- & $73\pdeg5 \pm 2\pdeg0$ & $23\pdeg5 \pm 1\pdeg1$\cr              
\noalign{\vskip 5pt\hrule\vskip 3pt}}}
\endPlancktablewide
\par
\endgroup
\end{table*}

\begin{figure}
\includegraphics[width=8.8cm]{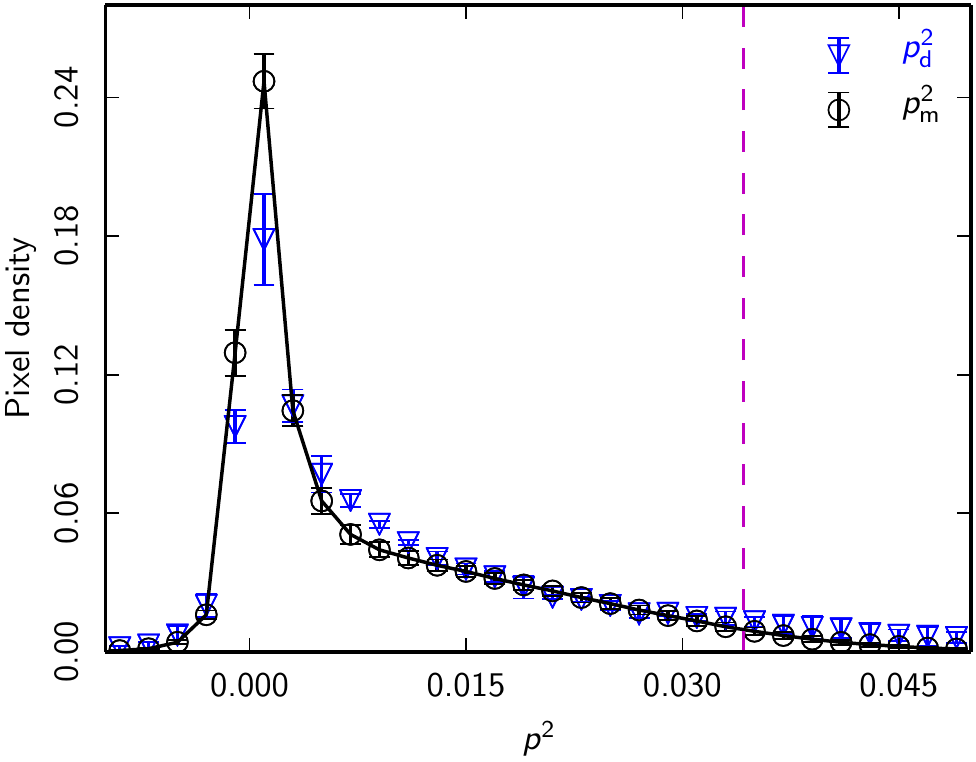} 
\caption{Normalized histograms of $p^2$ over SGC34 at 353\,GHz. \planck\ data, $p_{\rm d}^2$, blue inverted-triangles, with $1\sigma$ error bars 
as described in text. The pixel density on the y-axis is the number of pixels in each bin of $p^2$ divided by the total number of pixels over SGC34.
Dust model, $p_{\rm m}^2$, black circles connected by line, with $1\sigma$ error bars computed from 100 Monte-Carlo realizations. 
The vertical dashed line corresponds to the parameter $p_0=18.5\,\%$ ($p_0^2=0.034$) of this dust model. }
\label{fig:3.1.1}
\end{figure}

\subsection{Polarization fraction}
\label{sec:3.1.1}

The naive estimator of $p$, defined as $p = \sqrt{Q^2 + U^2}/I$, is a biased quantity. The bias in $p$ becomes dominant in the low signal-to-noise regime \citep{Simmons:1985}. \citet{Montier:2015} review the different algorithms to debias the $p$ estimate. To avoid the debiasing problem, we choose to work with the unbiased quantity $p_{\rm d}^2$  derived from a combination of independent subsets of the \planck\ data at 353\,GHz (see Equation~(12) of \citealt{planck2016-XLIV}):
\begin{equation}
p_{\rm d}^2  = \left < \frac{Q_{\rm d, 353}^{\rm s_1} \, Q_{\rm d, 353}^{\rm s_2}  \, + U_{\rm d, 353}^{\rm s_1} \, U_{\rm d, 353}^{\rm s_2}}{D_{353}^{2}} \right >\ ,
\label{eq:p2}
\end{equation}
where the subscript `${\rm d}$' refers to the observed polarized dust emission from the \planck\ 353\,GHz data (as distinct from `${\rm m}$' for the model below). 

Normalized histograms of $p_{\rm d}^2$ are computed over SGC34 from the cross-products of the three subsets of the \planck\ data where $(\rm s_1, s_2)$ = \{(HM1,HM2), (Y1,Y2), (DS1,DS2)\}.  Figure~\ref{fig:3.1.1} presents the mean of these three normalized histograms  (blue inverted-triangles). The two main sources of uncertainty that bias the normalized histogram of $p_{\rm d}^2$ are data systematics and the uncertainty of the zero-level of the Galactic dust emission. The former is computed per histogram bin using the standard deviation of $p_{\rm d}^2$ from the three subsets of the \planck\ data. The latter is
propagated by changing the dust intensity \Dmodel\ by $\pm\,3\KJysr$. The total $1\sigma$ error bar per histogram bin is the quadrature sum of these two uncertainties.

The mean, median, and maximum values of the dust polarization fraction for the three subsets of the \planck\ data are presented in Table~\ref{tab:1}. 
The maximum value of $p_{\rm d}$ at 1\deg\ resolution is consistent with the ones reported in \citet{planck2016-XLIV} over the high latitude southern Galactic cap and the \citet{planck2014-XIX} value at low and intermediate Galactic latitudes. We define a sky fraction $f(p_{\rm d} > 25\,\%)$ where the observed dust polarization fraction is larger than a given value of 25\,\% or $p_{\rm d}^2 > 0.0625$. 
In SGC34 this fraction is about 1\,\%. The small scatter between the results from the three different subsets shows that the \planck\ measurements are robust against any systematic effects.

The negative values of $p_{\rm d}^2$ computed from Eq.~(\ref{eq:p2}) arise from instrumental noise present in the three subsets of the \planck\ data. The width of the $p_{\rm d}^2$ normalized histogram is narrower than the width reported by \citet{planck2016-XLIV} over the high latitude southern Galactic cap. We were able to reproduce the broader distribution of $p_{\rm d}^2$ by adopting the same analysis region as in \citet{planck2016-XLIV}, showing the sensitivity to the choice of sky region.

\begin{figure}
\includegraphics[width=8.8cm]{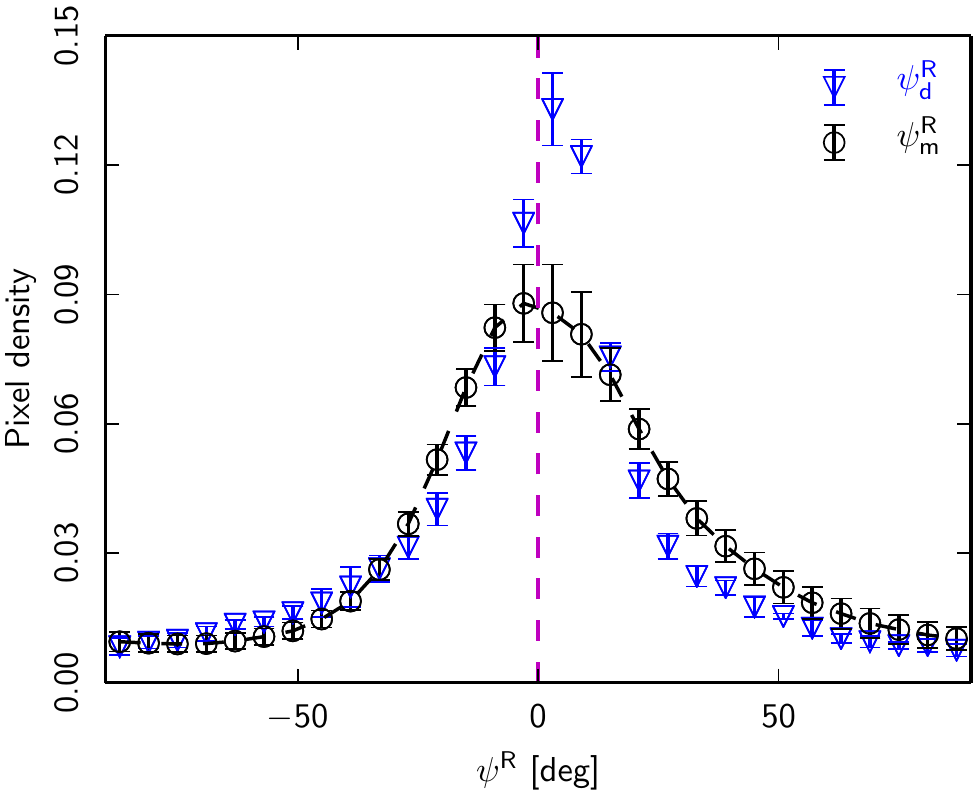}
\caption{Similar to Fig.~\ref{fig:3.1.1}, but for normalized histograms of $\psi^{\rm R}$, the polarization angle relative to the local polarization angle of the large-scale GMF.
The vertical dashed line corresponds to a model in which the turbulence is absent ($\Bt=0$ so that $\psi^{\rm R} = 0\deg$).}
\label{fig:3.1.1a}
\end{figure}

\subsection{Polarization angle}
\label{sec:3.1.2}

We use multiple subsets of the \planck\ polarization data at 353\,GHz to compute the normalized Stokes parameters $q_{\rm d, 353}^{\rm s} = Q_{\rm d, 353}^{\rm s}/D_{353}$ and $u_{\rm d, 353}^{\rm s}= U_{\rm d, 353}^{\rm s}/D_{353}$,
where $s$ = \{DS1, DS2, HM1, HM2, Y1, Y2\}. 
Using these data over SGC34, we fit the mean direction of the large-scale GMF, as defined by two coordinates $l_0$ and $b_0$, using model A of  \citet{planck2016-XLIV}. The best-fit values of $l_0$ and $b_0$ for each of the six independent subsets of the \planck\ data are quoted in Table~\ref{tab:1}. The mean values of $l_0$ and $b_0$ and their dispersions from the independent subsets are $73\pdeg5 \pm 2\pdeg0$ and $23\pdeg5\pm 1\pdeg1$, respectively. Our error bars are smaller than those (5\deg) reported by \citet{planck2016-XLIV} because the latter include the difference between fit values 
obtained with and without applying the bandpass mismatch (BPM) correction described in Sect. A.3 of \citet{planck2014-a09}. The mean values of $l_0$ and $b_0$ reported by \citet{planck2016-XLIV} were derived using maps with no correction applied. The difference of $3\pdeg5$ relative to our mean value of $l_0$ arises primarily from the BPM correction used in our analysis.

The best-fit values of $l_0$ and $b_0$ are used to derive a map of a mean-field polarization angle $\psi_0 (\hat{\vec{n}})$ (see \citealt{planck2016-XLIV} for the $Q$ and $U$ patterns associated with the mean field). We then rotate each sky pixel with respect to the new reference angle $\psi_0 (\hat{\vec{n}})$, using the relation
\begin{align}
&\left[\begin{array}{c}
Q_{\rm d, 353}^{\rm R} (\hat{\vec{n}}) \\
U_{\rm d, 353}^{\rm R} (\hat{\vec{n}}) \end{array}\right] = \nonumber \\
&\left(\begin{array}{ccc}
\cos{2 \psi_0} (\hat{\vec{n}}) & - \sin{2 \psi_0} (\hat{\vec{n}})  \\
\sin{2 \psi_0} (\hat{\vec{n}}) & \cos{2 \psi_0} (\hat{\vec{n}}) \end{array}\right)
 \left[\begin{array}{c}
Q_{\rm d, 353}^{\rm s}(\hat{\vec{n}}) \\
U_{\rm d, 353}^{\rm s} (\hat{\vec{n}}) \end{array}\right] \ ,
\end{align}
where $Q_{\rm d, 353}^{\rm R}$ and $U_{\rm d, 353}^{\rm R}$ are the new rotated Stokes parameters. The ``butterfly patterns'' caused by the large-scale GMF towards the southern Galactic cap \citep{planck2016-XLIV} are now removed from the dust polarization maps. Non-zero values of the polarization angle, $\psi_{\rm d}^{\rm R}$, derived from the $Q_{\rm d, 353}^{\rm R}$ and $U_{\rm d, 353}^{\rm  R}$  maps via
\begin{equation}
\psi_{\rm d}^{\rm R} (\hat{\vec{n}}) = \frac{1}{2} \tan^{-1} ( - U_{\rm d, 353}^{\rm R}(\hat{\vec{n}})  \, , \,Q_{\rm d, 353}^{\rm R} (\hat{\vec{n}}) ) \ ,    
\end{equation}
result from the dispersion of $\Bpos$ around the mean direction of the large-scale GMF. 
The minus sign in $U_{\rm d, 353}^{\rm R}$ is necessary to produce angles in the {\tt IAU} convention from \healpix\ format Stokes maps in the ``{\tt COSMO}" convention as used for the \planck\ data.

The mean normalized histogram of $\psi_{\rm d}^{\rm R}$ is computed from the multiple subsets of the \planck\ data and is presented in Fig.~\ref{fig:3.1.1a} (blue inverted-triangles). The $1\sigma$ error bars on  $\psi_{\rm d}^{\rm R}$ are derived from the standard deviation of the six independent measurements of the \planck\ data, as listed in Table~\ref{tab:1}. We fit a Gaussian to the normalized histogram of $\psi_{\rm d}^{\rm R}$ and find a $1\sigma$ dispersion of $15\pdeg0 \pm 0\pdeg4$ over SGC34. Our $1\sigma$ dispersion value of $\psi_{\rm d}^{\rm R}$ is slightly higher than the one found over the high latitude southern Galactic cap \citep{planck2016-XLIV}. This result is not unexpected because SGC34 extends up to $b \le -30\deg$ and adopting a mean direction of the large-scale GMF is only an approximation.

\begin{figure*}
\begin{tabular}{cc}
\includegraphics[width=8.8cm]{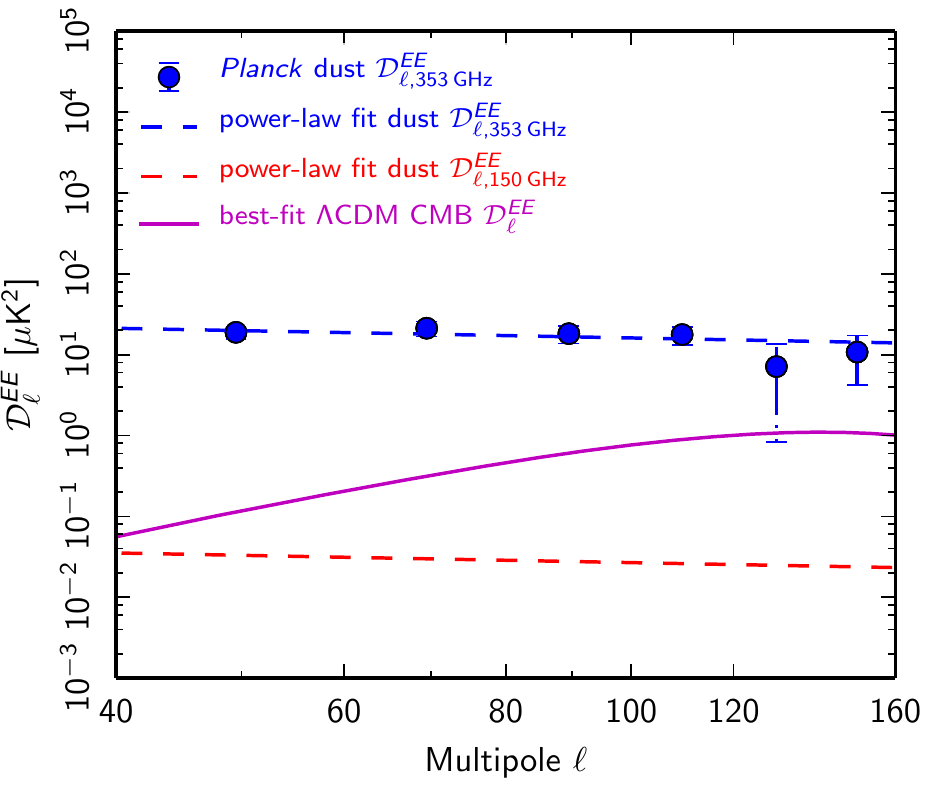} &
\includegraphics[width=8.8cm]{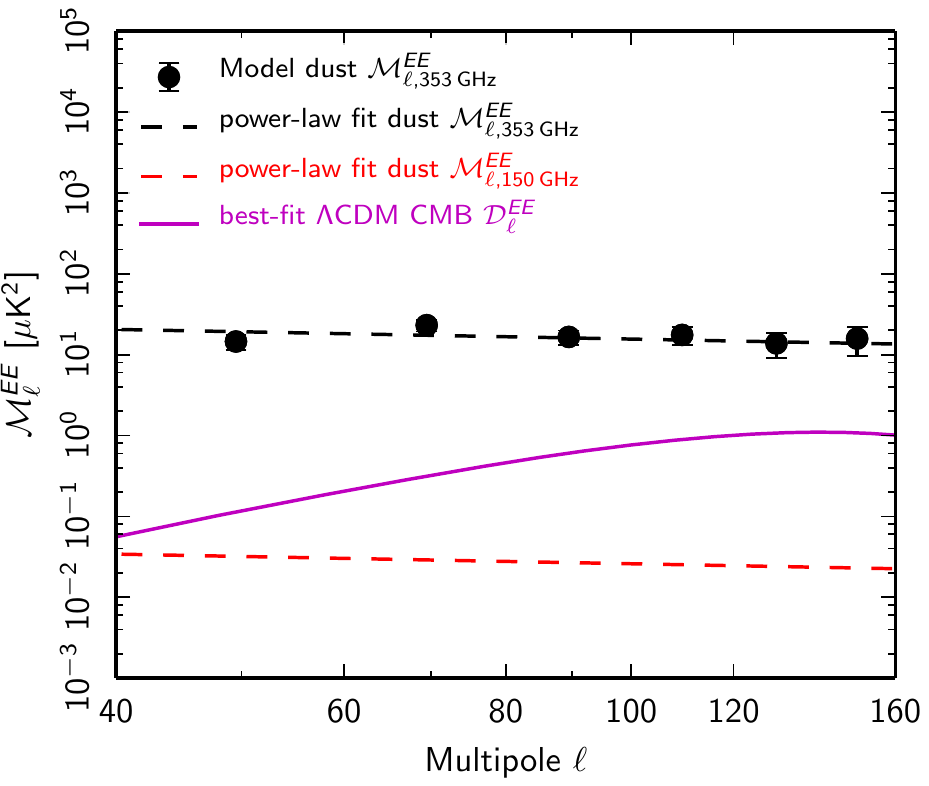}\\
\includegraphics[width=8.8cm]{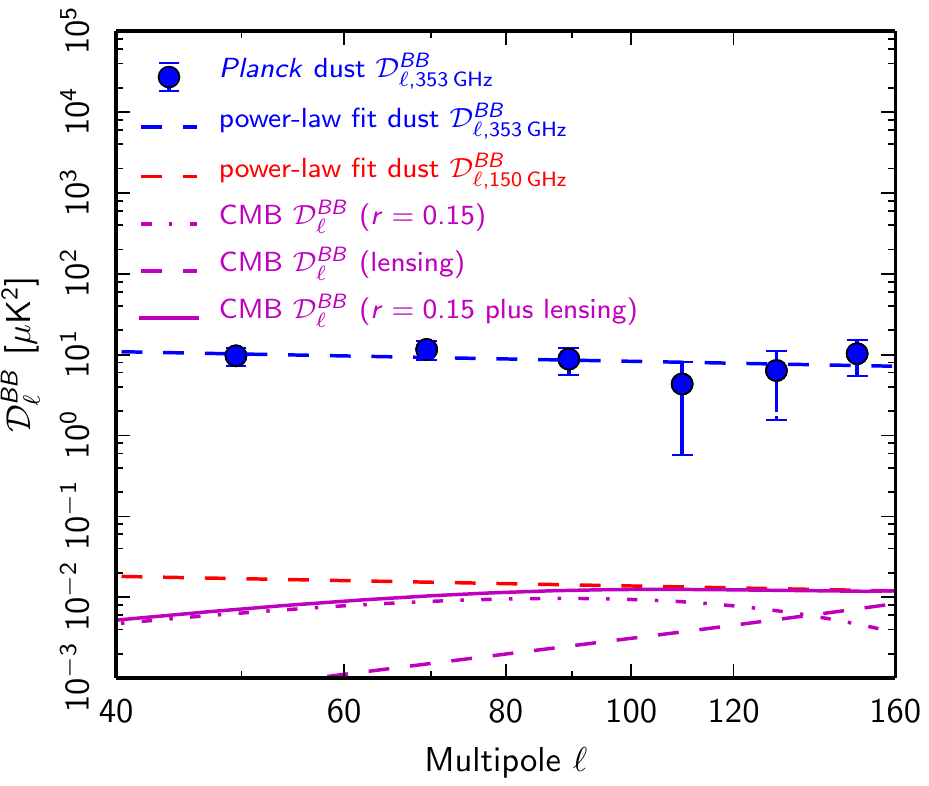} &
\includegraphics[width=8.8cm]{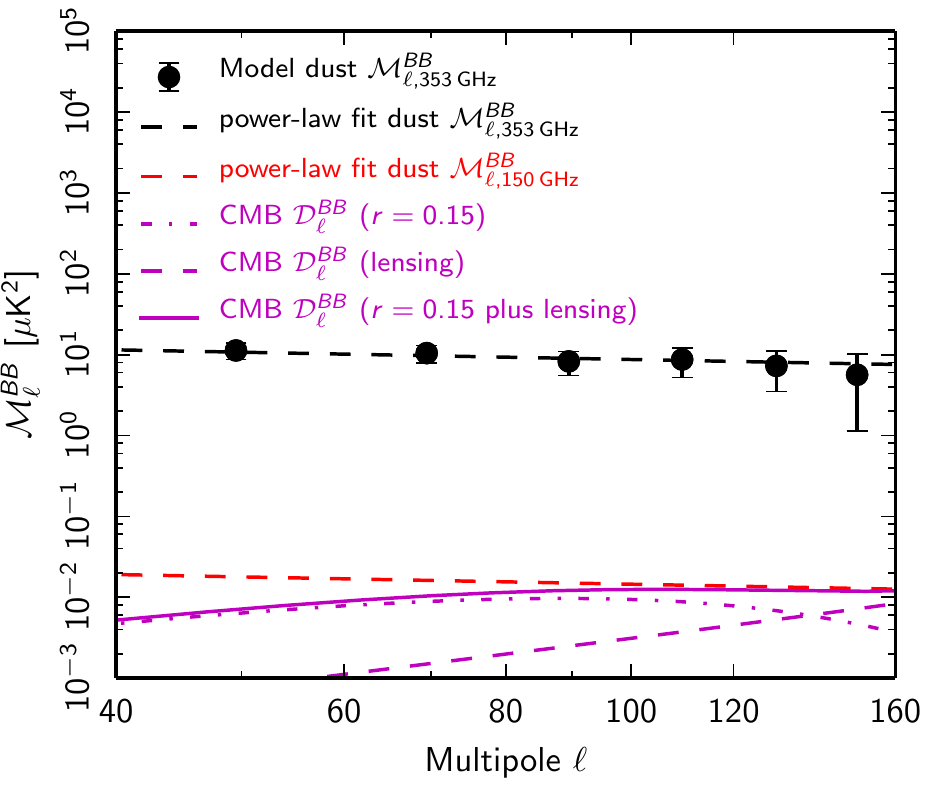}\\
\includegraphics[width=8.8cm]{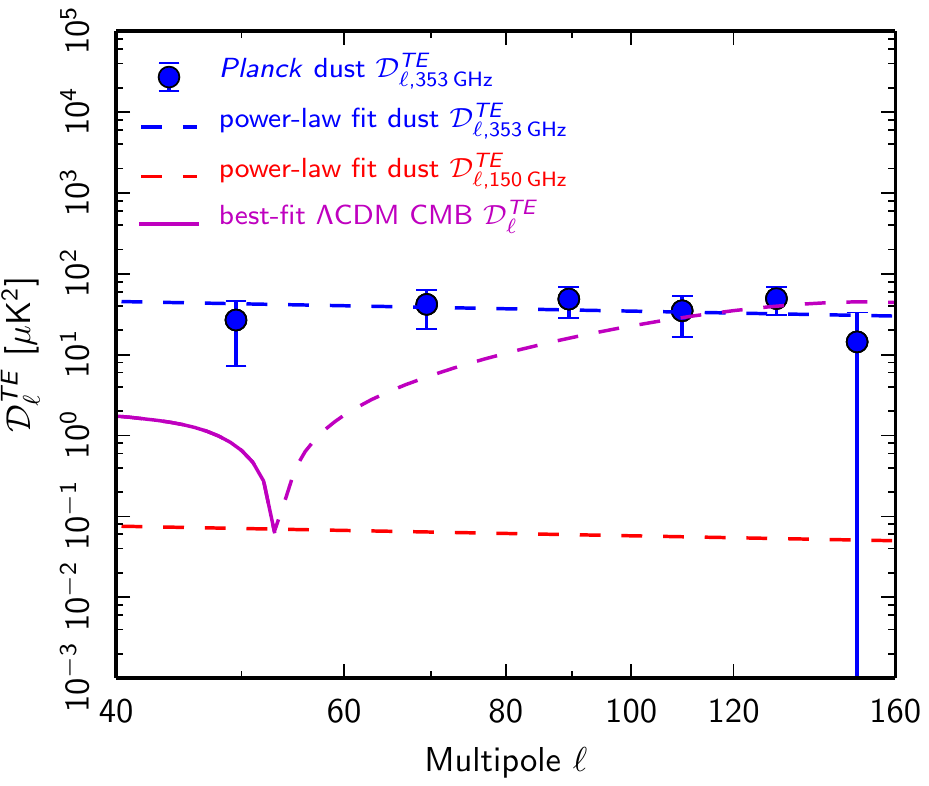} &
\includegraphics[width=8.8cm]{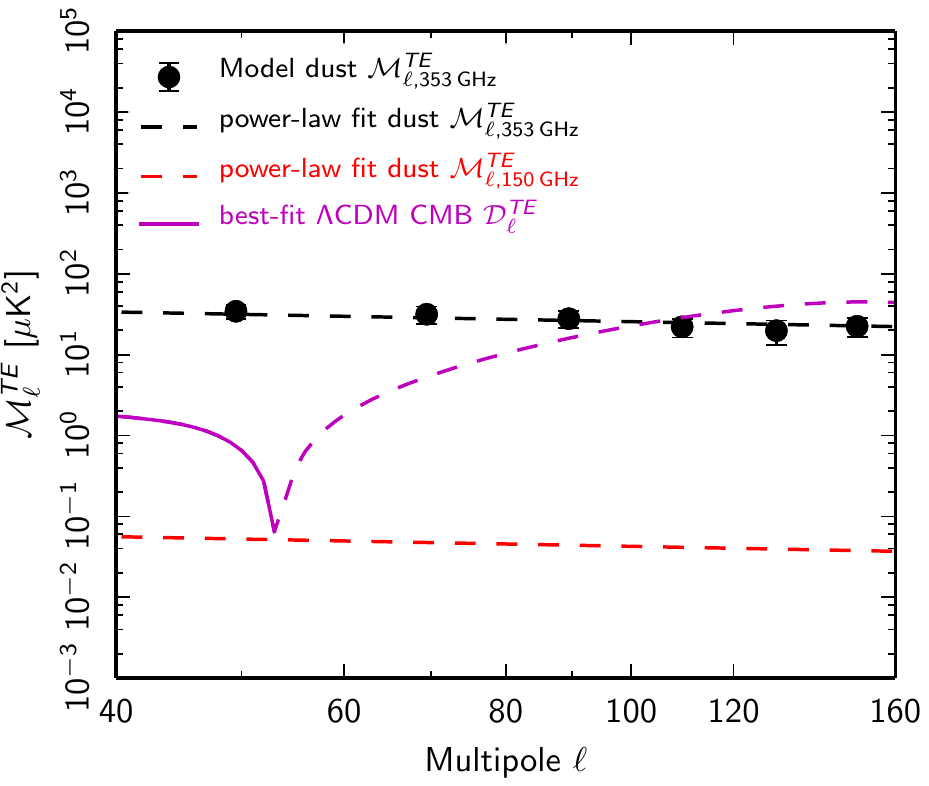}
\end{tabular}
\caption{\textit{Left column}: \planck\ 353\,GHz \dlee\  (\textit{top row}),  \dlbb\  (\textit{middle row}),  and  \dlte\  (\textit{bottom row}) power spectra in units of $\mukcmbsq$ computed over SGC34 using the subsets of the \planck\ data. See text re calculation of $1\sigma$ error bars. The best-fit power law assuming exponent $-2.3$ is plotted for each power spectrum (dashed blue line). For comparison, we also show the \planck\ 2015 best-fit $\Lambda$CDM expectation curves for the CMB signal \citep{planck2014-a15}, with the $TE$ expectation shown as dashed where it is negative.  Extrapolations of the dust power spectra to 150\,GHz (see text) are plotted as dashed red lines. \textit{Right column}: Similar to left panels, but for the dust model (Sect.~\ref{sec:results}).}
\label{fig:3.2}
\end{figure*}

\begin{table}[tb]
\begingroup
\newdimen\tblskip \tblskip=5pt
\caption{Fitted 353\,GHz dust power spectra of the \planck\ data and of the dust model over SGC34, for which the mean dust intensity $\langle \Dmodel \rangle$ is $54 \pm 3 \KJysr$.}
\label{table2}
\nointerlineskip
\vskip -3mm
\footnotesize
\setbox\tablebox=\vbox{
   \newdimen\digitwidth 
   \setbox0=\hbox{\rm 0} 
   \digitwidth=\wd0 
   \catcode`*=\active 
   \def*{\kern\digitwidth}
   \newdimen\signwidth 
   \setbox0=\hbox{+} 
   \signwidth=\wd0 
   \catcode`!=\active 
   \def!{\kern\signwidth}
\halign{
\hbox to 2.5cm{#\hfil}\tabskip 2.0em&
\hfil #\hfil &
\hfil #\hfil\tabskip=0pt\cr
\noalign{\doubleline}
 \omit Parameter \hfil & \planck\ 353\,GHz data & Dust model   \cr
\noalign{\doubleline}
$\alpha_{EE}$ & $-2.37\pm 0.29$ & $-2.02\pm 0.26$ \cr
$\alpha_{BB}$ & $-2.33\pm 0.42$ & $-2.48\pm 0.38$ \cr
$\alpha_{TE}$ & $-2.07\pm 0.58$ & $-2.48\pm 0.25$ \cr
\noalign{\doubleline}
Assuming a common power law exponent $\alpha_{XX} = -2.3$\cr
$\chi^2_{EE} (N_{\rm d.o.f.} = 6)$ & 2.78 & 5.17 \cr
$\chi^2_{BB} (N_{\rm d.o.f.} = 6)$ & 2.07 & 0.46\cr
$\chi^2_{TE} (N_{\rm d.o.f.} = 6)$ & 2.75 & 0.94 \cr
\noalign{\vskip 4pt\hrule\vskip 6pt}
$A_{EE}$ [$\mukcmbsq$] &  $17.21 \pm 1.82$  & $16.66 \pm 1.59$\cr
$A_{BB}$ [$\mukcmbsq$]  &  $8.85 \pm 1.36$  & $9.31 \pm 1.23$\cr
$A_{TE}$ [$\mukcmbsq$] &  $36.95 \pm 8.17$  &$27.37\pm 2.76$\cr
\noalign{\vskip 4pt\hrule\vskip 6pt}
$\langle A_{BB} /A_{EE} \rangle$ &$0.51\pm 0.10$ & $0.56\pm 0.09$\cr
$\langle A_{TE} /A_{EE} \rangle$ &$2.18\pm 0.54$ &$1.64\pm 0.23$\cr
\noalign{\vskip 5pt\hrule\vskip 3pt}}}
\endPlancktablewide
\par
\endgroup
\end{table}

\subsection{Polarized dust power spectra}
\label{sec:3.1.3}

We extend the analysis by \citet{planck2014-XXX} of the region LR24 ($f_{\rm sky}^{\rm eff}$\,=\,0.24 or 24\,\% of the total sky) to the smaller lower column density selected area SGC34 within it ($f_{\rm sky}^{\rm eff}$\,=\,0.085). We compute the polarized dust power spectra \dlee, \dlbb, and \dlte\ by cross-correlating the subsets of the \planck\ data at 353\,GHz  in the multipole range $40 < \ell < 160$ over SGC34. We use the same \Dmodel\ map for all subsets of the \planck\ data. The binning of the measured power spectra in six multipole bins matches that used in \citet{planck2014-XXX}. We take the mean power spectra computed from the cross-spectra of DetSets (DS1$\times$DS2), HalfMissions (HM1$\times$HM2), and Years (Y1$\times$Y2). The mean dust \dlee\ is corrected for the CMB contribution using the \planck\ best-fit $\Lambda$CDM model at the power spectrum level, while the mean \dlbb\ and \dlte\ are kept unaltered. Figure~\ref{fig:3.2} shows that the contribution of the CMB $BB$ spectra (for $r=0.15$ with lensing contribution) is negligibly small as compared to the mean dust spectra \dlbb\ over SGC34. The mean \dlte\ is unbiased because we use \Dmodel\ as a  dust-only Stokes $I$ map. However, the unsubtracted CMB $E$-modes bias the $1\sigma$ error of the dust \dlte\ power spectrum by a factor $\sqrt{[ \dltt  (\text{dust}) \times  \dlee (\text{CMB}) ]/\nu_{\ell}}$, where $\nu_{\ell}$ is associated with the effective number of degrees of freedom \citep{Tristram:2005}.

The mean dust power spectra \dlee, \dlbb, and \dlte\ of the \planck\ 353\,GHz data are presented in the left column of Fig.~\ref{fig:3.2}.   The $1\sigma$ statistical uncertainties from noise (solid error bar in the  left column of Fig.~\ref{fig:3.2}) are computed from the analytical approximation implemented in \xpol. The $1\sigma$ systematic uncertainties are computed using the standard deviation of the dust polarization power spectra from the three subsets of the \planck\ data (as listed in Table~\ref{tab:1}). The total $1\sigma$ error bar per $\ell$ bin (dashed error bar in the  left column of Fig.~\ref{fig:3.2}) is the quadrature sum of the two. The measured \planck\ 353\,GHz polarized dust power spectra \dlee, \dlbb, and \dlte\ are well described by power laws in multipole space, consistent with the one measured at intermediate and high Galactic latitudes \citep{planck2014-XXX}. Similar to the analysis in \citet{planck2014-XXX}, we fit the power spectra with the power-law model, $\dlxx = A_{XX} (\ell/80)^{\alpha_{XX}+2}$, where $A_{XX}$ is the best-fit amplitude at $\ell =80$, $\alpha_{XX}$ is the best-fit slope, and $XX=\{EE, BB, TE\}$. We restrict the fit to six band-powers in the range $40 < \ell < 160$ so that we can compare the \planck\ data directly with the low-resolution dust model derived using the GASS \hi\ data.  The exponents $\alpha_{XX}$ of the unconstrained fits are quoted in the top row of Table~\ref{table2}. Next, we perform the fit with a fixed exponent $-2.3$ for the appropriate $\alpha_{XX}$.  The resulting power laws are shown in the left column of Fig.~ \ref{fig:3.2} and the values of $\chi^2$ (with number of degrees of freedom, $N_{\rm d.o.f}=6$) and the best-fit values of $A_{XX}$ are presented in Table~\ref{table2}.  

The ratio of the dust $B$- and $E$-mode power amplitudes, roughly equal to a half (Table~\ref{table2}), is maintained from LR24 to SGC34. The measured amplitude $A_{BB} = 8.85\,\mukcmbsq$ at $\ell=80$ is about $\sim35\%$ lower than the value expected from the empirical power-law $A_{BB} \propto \langle \Dmodel \rangle^{1.9}$ \citep{planck2014-XXX}  applied to SGC34. For future work, it will be important to understand the amplitude and variations of the dust $B$-mode signal in these low column density regions.

We find a significant positive $TE$ correlation, \dlte, over SGC34; the amplitude $A^{TE}$ normalized at $\ell=80$ has more than $4\,\sigma$ significance when the multipoles between 40 and 160 are combined in this way. Similarly, a positive $TE$ correlation was reported over LR24 (see Figure B.1 of \citealt{planck2014-XXX}). The non-zero positive $TE$ correlation is a direct consequence of the correlation between the dust intensity structures and the orientation of the magnetic field. \citet{planck2015-XXXVIII} report that oriented stacking of the dust $E$-map over $T$ peaks of interstellar filaments identified in the dust intensity map picks up a positive $TE$ correlation (see their Figure 10).

For comparison in Fig.~\ref{fig:3.2} we show the \planck\ 2015 best-fit $\Lambda$CDM CMB \dlee,  \dlbb\ (primordial tensor-to-scalar ratio $r=0.15$ plus lensing contribution), and \dlte\ expectation curves  \citep{planck2014-a15}. Then assuming that the dust spectral energy distribution (SED) follows a MBB spectrum with a mean dust spectral index $1.59\pm 0.02$ and a mean dust temperature of 19.6\,K \citep{planck2014-XXII}, we extrapolate the best-fit dust power spectra from 353\,GHz to 150\,GHz (scaling factor $s_{150/353}=0.0408^2$) in order to show the level of the dust polarization signal (red dashed lines in Fig.~\ref{fig:3.2}) relative to the CMB signal over SGC34.

\section{Model framework}
\label{sec:methodology}

In this section we extend the formalism developed in \citet{planck2016-XLIV} to simulate the dust polarization over SGC34 using a phenomenological description of the magnetic field and the GASS \hi\ emission data. We introduce two main ingredients in the dust model: (1) fluctuations in the GMF orientation along the LOS and (2) alignment of the CNM structures with \Bpos. Based on the filament study by \citet{planck2015-XXXVIII}, we anticipate that alignment of the CNM structures with \Bpos\ will account for the positive dust $TE$ correlation and the $\dlte/\dlee$ ratio over SGC34.

\subsection{Formalism}
\label{sec:4.1.1}

We use the general framework introduced by \citet{planck2014-XX} to model the dust Stokes parameters, $I$, $Q$, and $U$, for optically thin regions at a frequency $\nu$:
\begin{align}
I_{\rm m, \nu}   (\hat{\vec{n}}) &= \int  \left [1- p_0 \left ( \cos^2 \gamma^i(\hat{\vec{n}}) - \frac{2}{3} \right ) \right ]  S_{\nu}^i (\hat{\vec{n}}) \ d\tau_{\nu}  \nonumber \\
Q_{\rm m, \nu} (\hat{\vec{n}}) &= \int  p_0 \cos^2 \gamma^i(\hat{\vec{n}}) \cos 2\psi^i(\hat{\vec{n}})  S_{\nu}^i (\hat{\vec{n}}) \ d\tau_{\nu}  \label{eq:4.1}\\
U_{\rm m, \nu} (\hat{\vec{n}}) &= - \int  p_0 \cos^2 \gamma^i(\hat{\vec{n}}) \sin 2\psi^i(\hat{\vec{n}}) S_{\nu}^i (\hat{\vec{n}}) \ d\tau_{\nu}  \ , \nonumber
\end{align}
where (see Appendix B of \citealp{planck2014-XX} for details) $\hat{\vec{n}}$ is the direction vector, $S_{\nu}$ is the source function given by the relation $S_{\nu}= n_{\rm d} B_{\nu} (T) C_{\rm avg}$, $\tau_{\nu}$ is the optical depth, $p_0 = p_{\rm dust} R$ is the product of the ``intrinsic dust polarization fraction" ($p_{\rm dust}$) and the Rayleigh reduction factor ($R$, related to the alignment of dust grains with respect to the GMF), $\psi$ is the polarization angle measured from Galactic North, and $\gamma$ is the angle between the local magnetic field and the POS. Again, like in Sect.~\ref{sec:3.1.2}, the minus sign in $U_{\rm m, \nu}$ is necessary to work from the angle map $\psi$ given in the {\tt IAU} convention to produce \healpix\ format Stokes maps in the {\tt COSMO} convention.  We start with the same dust Stokes parameters equations as in the \citet{Vansyngel:2016} parametric dust model.

Following the approach in \citet{planck2016-XLIV}, we replace the integration along the LOS over SGC34 by the sum over a finite number of layers with different polarization properties. These layers are a phenomenological means of accounting for the effects of fluctuations in the GMF orientation along the LOS, the first main ingredient of the dust model.  We interpret these layers along the LOS as three distinct phases of the ISM. The Gaussian decomposition of total \NHI\ into the CNM, UNM, and WNM components based on their line-widths (or velocity dispersion) is described in detail in Sect.~\ref{sec:2.3}.

We can replace the integral of the source function over each layer directly with the product of the column density maps of the three \hi\ components and their mean dust emissivities. Within these astrophysical approximations, we can modify Eq.~\eqref{eq:4.1} to be
\begin{align}
I_{\rm m, \nu} (\hat{\vec{n}}) &=  ~\sum_{i=1}^{3}  \left [1- p_0 \left ( \cos^2 \gamma^i(\hat{\vec{n}}) - \frac{2}{3} \right ) \right ] \langle \epsilon_{\nu}^i \rangle  \NHIi (\hat{\vec{n}})  \nonumber \\
Q_{\rm m, \nu}(\hat{\vec{n}}) &= ~ \sum_{i=1}^{3}    p_0  \cos^2 \gamma^i(\hat{\vec{n}})  \cos 2\psi^i(\hat{\vec{n}})  \langle \epsilon_{\nu}^i \rangle  \NHIi (\hat{\vec{n}})  \label{eq:4.2}  \\
U_{\rm m, \nu}(\hat{\vec{n}}) &= - \sum_{i=1}^{3}   p_0  \cos^2 \gamma^i(\hat{\vec{n}})  \sin 2\psi^i(\hat{\vec{n}})  \langle \epsilon_{\nu}^i \rangle  \NHIi (\hat{\vec{n}})  \ ,\nonumber
\end{align}
where $\langle \epsilon_{\nu}^i \rangle$ is the mean dust emissivity for each of the three \hi\ components. 

In general, the emissivity is a function not only of frequency, but also of sky position $\hat{\vec{n}}$ and different \hi\ phases. However, in SGC34 the overall emissivity \Dmodel/\NHI\ is quite uniform (see Sect.~\ref{choiceparams}) and from a simultaneous fit of the \Dmodel\ map to all three component \NHI\ maps we find little evidence, for our decomposition, that the emissivities of the individual components are different than the overall mean.   

Similarly, the effective degree of alignment $p_0=p_{\rm dust} R$ could be a function of frequency and different for different \hi\ components.  Equation~\eqref{eq:4.2} also tells us that the product $p_0$ is degenerate with $ \langle \epsilon_{\nu} \rangle$ for the dust Stokes $Q_{\rm m, \nu}$ and $U_{\rm m, \nu}$ parameters. Such effects could make the source functions in the layers even more different from one another.

In this paper we explore how models based on simplifying assumptions, namely a constant $p_0$ and a frequency dependent $\langle \epsilon_{\nu} \rangle$ common to the three \hi\ components, might despite the reduced flexibility nevertheless provide a good description of the data.

This use of the GASS \NHI\ data can be contrasted to the approach of \citet{planck2016-XLIV} and \citet{Vansyngel:2016}, where the source functions are assumed to be the same in each layer.  Furthermore, \citet{planck2016-XLIV} choose to ignore the correction term involving $p_0$ in the integrand for $I_{\rm m, \nu} (\hat{\vec{n}})$.

This phenomenological dust model based on the GASS \hi\ data is not unique. It is based on several astrophysical assumptions as noted in this section. Within the same general framework, we could contemplate replacing the \hi\ emission with three-dimensional (3D) extinction maps, though no reliable maps are currently available over the SGC34 region \citep{Green:2015}. Furthermore, the \hi\ emission traces the temperature or density structure of the diffuse ISM, whereas the dust extinction traces only the dust column density. Thus, using the \hi\ emission to model the dust polarization is an interesting approach even if we had reliable 3D extinction maps toward the southern Galactic cap.

\subsection{Structure of the Galactic magnetic field}
\label{sec:4.1.2}

The Galactic magnetic field, \BT, is expressed as a vector sum of a mean large-scale (ordered, \Bo) and a turbulent (random, \Bt) component \citep{Jaffe:2010}
\begin{align}
\BT(\hat{\vec{n}}) &= \Bo(\hat{\vec{n}}) +  \Bt(\hat{\vec{n}}) \nonumber  \\
&=|\Bo| ( \uBo(\hat{\vec{n}}) +  f_{\rm M} \uBt(\hat{\vec{n}})) \ , \label{eq:4.3}
\end{align}
where $f_{\rm M}$ is the standard deviation of the relative amplitude of the turbulent component $|\Bt |$ with respect to the mean large-scale $|\Bo |$.  The butterfly patterns seen in the orthographic projections of the \planck\ polarization maps, $Q_{\rm d, 353}$ and $U_{\rm d, 353}$, centred at  ($l,b)=(0\deg, -90\deg$) are well fitted by a mean direction of the large-scale GMF over the southern Galactic cap \citep{planck2016-XLIV}. We assume that the mean direction of the large-scale GMF is still a good approximation to fit the butterfly patterns seen in the \planck\ Stokes $Q_{\rm d, 353}$ and $U_{\rm d, 353}$ maps over SGC34 (Fig.~\ref{fig:4.2.1}). The unit vector \uBo\ is defined as $\uBo = (\cos l_0 \cos b_0, \sin l_0 \cos b_0, \sin b_0)$. 

We model \Bt\ with a Gaussian realization on the sky with an underlying power spectrum, $C_{\ell} \propto  \ell^{\alpha_{\rm M}}$, for $\ell \ge 2$ \citep{planck2016-XLIV}. The correlated patterns of \Bt\ over the sky are needed to account for the correlated patterns of dust $p$ and $\psi$ seen in \planck\ 353\,GHz data \citep{planck2014-XIX}. Following \citet{Fauvet:2012}, we only consider isotropic turbulence in our analysis. We produce a spatial distribution of \Bt\ in the $x$, $y$, and $z$ direction from the above power spectrum on the celestial sphere of \healpix\ resolution \Nside. We add \Bt\ to \Bo\ with a relative strength $f_{\rm M}$ as given in Eq.~\ref{eq:4.3}.

\subsection{Alignment of the CNM structures with the magnetic field}
\label{sec:4.1.3}

We use the algebra given in Section~4.1 of \citet{planck2016-XLIV} to compute $\cos^2\,\gamma^i$ for the three \hi\ components. For the UNM and WNM components, we assume that there is no $TE$ correlation and the angles $\psi^{\rm u}$ and $\psi^{\rm w}$ are computed directly from the total \BT\ again using the algebra by \citet{planck2016-XLIV}. For the CNM we follow a quite different approach to compute $\psi^{\rm c}$. 

We introduce a positive $TE$ correlation for the CNM component through the polarization angle $\psi^{\rm c}$. The alignment between the CNM structures and \Bpos\ is introduced  to test whether it fully accounts for the observed $E$-$B$ power asymmetry or not. To do that, we assume that the masked $\NHI^{\rm c}$ map is a pure $E$-mode map and define spin-2 maps $Q_{\rm T}$ and $U_{\rm T}$ as
\begin{equation}
(Q_{\rm T} \pm iU_{\rm T})(\hat{\vec{n}}) = \sum_{\ell=2}^{\infty} \sum_{m=-\ell}^{\ell} a_{\ell m}^{\rm c} \ {}_{\pm 2}Y_{\ell m}(\hat{\vec{n}})  \ ,
\end{equation}
where $a_{\ell m}^{\rm c}$ are the harmonic coefficients of the masked $\NHI^{\rm c}$ map. In the flat-sky limit, $Q_{\rm T} \simeq (\partial_{\rm x}^2 - \partial_{\rm y}^2) \nabla^{-2} \NHI^{\rm c}$ and $U_{\rm T} \simeq -2 \partial_{\rm x}\partial_{\rm y} \nabla^{-2} \NHI^{\rm c}$ , where $\nabla^{-2}$ is the inverse Laplacian operator \citep{Bowyer:2011}. We compute the angle $\psi^{\rm c}$ using the relation, $\psi^{\rm c} = (1/2) \tan ^{-1} (-U_{\rm T}, Q_{\rm T})$ (see \citealt{planck2014-a18} for details). The polarization angle $\psi^{\rm c}$ is the same as the angle between the major axis defined by a local quadrature expansion of the $\NHI^{\rm c}$ map and the horizontal axis. This idea of aligning perfectly the CNM structures with \Bpos\ is motivated from the \hi\ studies by \citet{Clark:2015}, \citet{Martin:2015}, and \citet{Kalberla:2016}, and \planck\ dust polarization studies by \citet{planck2014-XXXII} and \citet{planck2015-XXXVIII}. This is the second main ingredient of the dust model.

\subsection{Model parameters}
\label{sec:4.4}

The dust model has the following parameters: $\langle \epsilon_{\nu} \rangle$, the mean dust emissivity per \hi\ component at frequency $\nu$;  $p_0$, the normalization factor to match the observed dust polarization fraction over the southern Galactic cap;  $l_0$ and $b_0$, describing the mean direction of \Bo\ in the region of the southern Galactic cap; $f_{\rm M}$, the relative strength of \Bt\ with respect to \Bo\ for the different components; $\alpha_{\rm M}$, the spectral index of \Bt\ power spectrum for the different components; and $\sigma_{\rm c}$ and $\sigma_{\rm u}$ to separate the total \NHI\ into the CNM, UNM, and WNM components (phases).
 
We treat the three \hi\ components with different density structures and magnetic field orientations as three independent layers. Because we include alignment of the CNM structures with \Bpos, we treat the CNM polarization layer separately from the WNM and UNM polarization layers. The value of $f_{\rm M}^{\rm c}$ for the CNM component is different than the common $f_{\rm M}^{\rm u/w}$ for the UNM and WNM components. For the same reason, the spectral index of the turbulent power spectrum $\alpha_{\rm M}^{\rm c}$ for the CNM component could be different than the common $\alpha_{\rm M}^{\rm u/w}$ for the UNM and WNM components, but we take them to be the same, 
simplifying the model and reducing the parameter space in our analysis.
We also use the same values of $\langle \epsilon_{\nu} \rangle$ and $p_0$ for each layer.  This gives a total of nine parameters.

In this analysis, we do not optimize the nine parameters of the dust model explicitly. Instead, we determine five parameters based on various astrophysical constraints and optimize only $\langle \epsilon_{\nu} \rangle$, $p_0$, $f_{\rm M}^{\rm u/w}$, and $\sigma_{\rm c}$ simultaneously (Sect.~\ref{choiceparams}).  Our main goal is to test quantitatively whether the fluctuations in the GMF orientation along the LOS plus the preferential alignment of the CNM structures with \Bpos\ can account simultaneously for both the observed dust polarization power spectra and the normalized histograms of $p^2$ and $\psi^{\rm R}$ over SGC34.

We review here the relationship between the nine dust model parameters and the various dust observables.

\begin{itemize}

\item The emissivity $\langle \epsilon_{\nu} \rangle$ simply converts the observed \NHI\ ($10^{20} \cm2$) to the \planck\ intensity units (in $\mukcmb$) \citep{planck2013-XVII}.   This parameter is optimized in Sect.~\ref{choiceparams}, subject to a strong prior obtained by correlating the map \Dmodel\ with the \NHI\ map.
\item The parameter $p_0$ is constrained by the observed distribution of $p_{\rm d}^2$ at 353\,GHz and the amplitude of the dust polarization power spectra \dlee, \dlbb, and \dlte. It is optimized using the latter in Sect.~\ref{choiceparams}.
\item The two model parameters $l_0$ and $b_0$ determine the orientation of the butterfly patterns seen in the orthographic projections of the Planck polarization maps, $Q_{\rm d, 353}$ and $U_{\rm d, 353}$ \citep{planck2016-XLIV}.  They are found by fitting these data (Sect.~\ref{sec:3.1.2}).
\item The parameter $f_{\rm M}$ for each polarization layer is related to the dispersion of $\psi_{\rm d}^{\rm R}$ and $p_{\rm d}^2$ (the higher the strength of the turbulent field with respect to the ordered field, the higher the dispersion of $\psi_{\rm d}^{\rm R}$ and $p_{\rm d}^2$) and to the amplitude of the dust polarization power spectra.
Our prescription for the alignment of CNM structure and \Bpos\ subject to statistical constraints from polarization data determines $f_{\rm M}^{\rm c}$ (Sect.~\ref{sec:cnm_alignment}); $f_{\rm M}^{\rm u/w}$ is optimized using the dust polarization band powers in Sect.~\ref{choiceparams}.
\item The parameter $\alpha_{\rm M}^{\rm c}$ is constrained by the alignment of the CNM structures with \Bpos\  (Sect.~\ref{sec:cnm_alignment}). 
\item  The parameter $\sigma_{\rm c}$ controls the fraction of the \hi\ in the CNM and hence sets the amount of $E$-$B$ power asymmetry and the $\dlte/\dlee$ ratio through the alignment of the CNM structures with \Bpos.  It is optimized in Sect.~\ref{choiceparams}.
Our dust model is insensitive to the precise value of the parameter $\sigma_{\rm u}$, which is used to separate the UNM and WNM into distinct layers. 

\end{itemize}

To break the degeneracy between different dust model parameters and obtain a consistent and robust model, we need to make use of dust polarization observables in pixel space as well as in harmonic space.

\section{Dust sky simulations}
\label{sec:model_params}

We simulate the dust intensity and polarization maps at only a single \planck\ frequency 353\,GHz. It would be straightforward to extrapolate the dust model to other microwave frequencies using a MBB spectrum of the dust emission \citep{planck2013-XVII, planck2014-XXII, planck2014-XXX} or replacing the value of $\langle\epsilon_{\nu} \rangle$ from \citet{planck2013-XVII}. However, such a simple-minded approach would not generate any decorrelation of the dust polarization pattern between different frequencies, as reported in \citet{planck2016-L}.

\begin{figure}
\begin{tabular}{c}
\includegraphics[width=8.8cm]{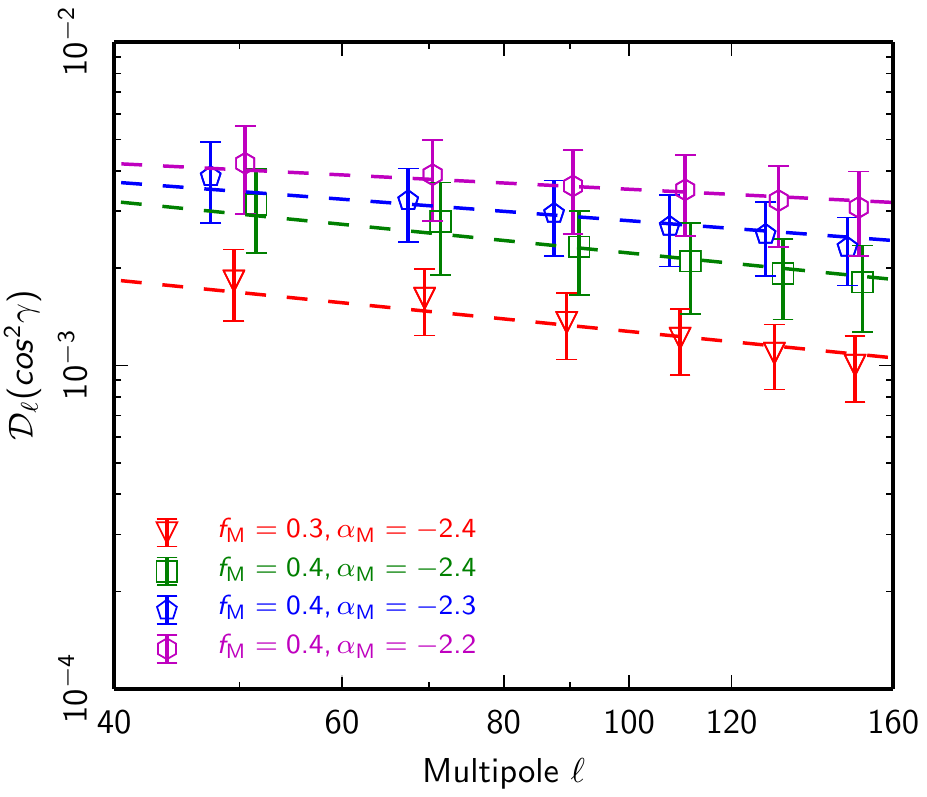} \\
\includegraphics[width=8.8cm]{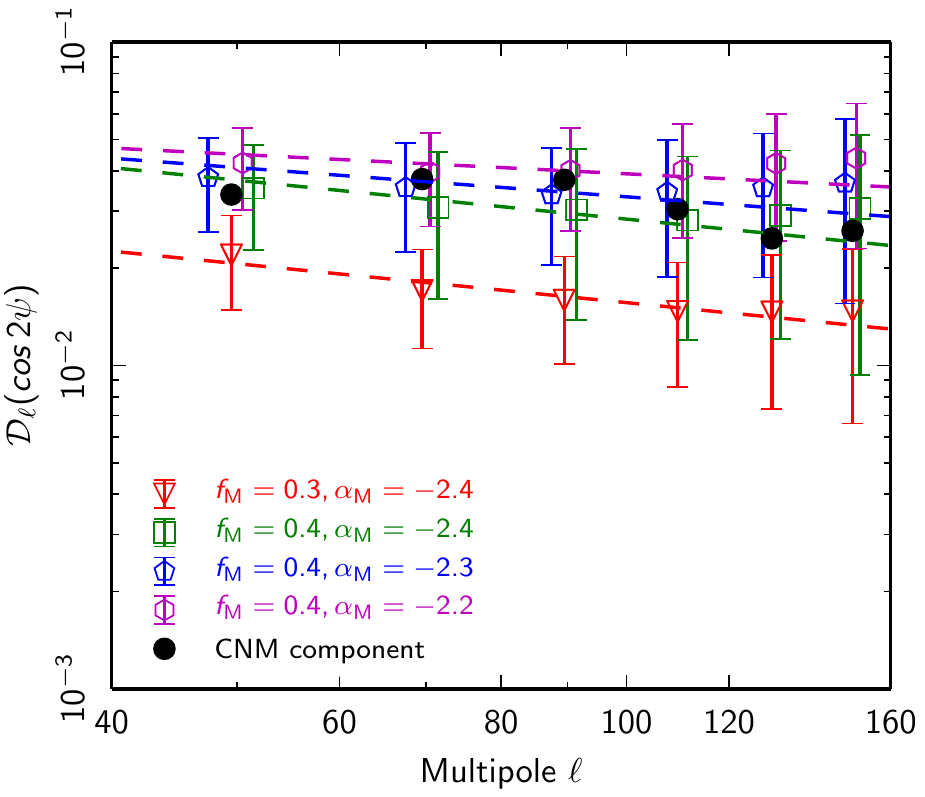}
\end{tabular}
\caption{Mean power spectra of $\cos^2 \gamma$ (\textit{top panel}) and $\cos 2\psi$ (\textit{bottom panel}) over SGC34 for a set of $\alpha_{\rm M}$ and $f_{\rm M}$ parameters in the multipole range $40< \ell< 160$. The $1\sigma$ error bars are computed from the standard deviations of the 100 Monte-Carlo realizations of the dust model. The alignment of the CNM structures with \Bpos\ (see Sect.~\ref{sec:cnm_alignment} for details) constrains the value of $\alpha_{\rm M}^{\rm c}$ and $f_{\rm M}^{\rm c}$ for the CNM component.}
\label{fig:4.3.1}
\end{figure}

\subsection{Constraining the turbulent GMF for the CNM component}
\label{sec:cnm_alignment}

As discussed in Sect.~\ref{sec:4.1.3}, we adopted an alternative approach to compute the polarization angle $\psi^{\rm c}$ for the CNM component. At first sight, it might look physically inconsistent and therefore a main source of bias. However, we now show that the original and alternative approaches lead to statistically similar results and so can be compared in order to constrain the parameters of \Bt, that is $\alpha_{\rm M}^{\rm c}$ and $f_{\rm M}^{\rm c}$, for the CNM component.

To do this we first adopt the original approach from \citet{planck2016-XLIV} (and used for the UNM and CNM components).  We generate Gaussian realizations of \Bt\ for a set of $(\alpha_{\rm M}, f_{\rm M})$ values. We assume a fixed mean direction of \Bo\ as $(l_0, b_0) = (73\pdeg5, 23\pdeg5)$ (Table~\ref{tab:1}) and add to it \Bt\ to derive a total \BT. We use the algebra given in Section~4.1 of \citet{planck2016-XLIV} to compute the two quantities $\cos^2\,\gamma$ and $\cos 2\psi$ for different values of $\alpha_{\rm M}$ and $f_{\rm M}$. Because the polarization angles $\gamma$ and $\psi$ are circular quantities, we choose to work with quantities like $\cos^2\,\gamma$ and $\cos 2\psi$ as they appear in the Stokes $Q$ and $U$ parameters (Eq.~\ref{eq:4.2}). We then apply the \xpol\ routine to compute the angular power spectra of these two quantities for each sky realization over SGC34. The mean power spectra of  $\cos^2\,\gamma$ and $\cos 2\psi$ are computed by averaging 100 Monte-Carlo realizations and are presented in Fig.~\ref{fig:4.3.1}. The power spectra of the two polarization angles  are well described by power-laws in multipole space, ${\cal D}_{\ell} \propto \ell^{\alpha_{\rm M} + 2 }$, where $\alpha_{\rm M}$ is the same as the spectral index of \Bt\ power spectrum. The amplitudes and slopes of these two quantities are related systematically to the values of $\alpha_{\rm M}$ and $f_{\rm M}$. The higher the $f_{\rm M}$ value, the higher is the amplitude of the power spectra in Fig.~\ref{fig:4.3.1}. The best-fit slope of the power spectrum over the multipole range $40 < \ell < 160$ is the same as the spectral index of the \Bt\ power spectrum for relevant $f_{\rm M}$ values. 

For each Monte Carlo realization we also calculated $\psi^{\rm c}$ for the CNM component by the alternative approach and from that computed the power spectra of $\cos 2\psi$ for the CNM component over SGC34. By comparing these results to the spectra from the first \citet{planck2016-XLIV} approach, as in Fig.~\ref{fig:4.3.1} (bottom panel), we can constrain the values of $\alpha_{\rm M}^{\rm c}$ and $f_{\rm M}^{\rm c}$. The typical values of $\alpha_{\rm M}^{\rm c}$ and $f_{\rm M}^{\rm c}$ are $-2.4$ and 0.4, respectively.  One needs a Monte-Carlo approach to put realistic error bars on the $\alpha_{\rm M}^{\rm c}$ and $f_{\rm M}^{\rm c}$ parameters, which is beyond the scope of this paper. However, we do note that the value of $\alpha_{\rm M}^{\rm c}$ is close to the exponent of the angular power spectrum of the $\NHI^{\rm c}$ map computed over SGC34, reflecting the correlation between the structure of the intensity map and the orientation of the field.

We stress that the values of $\alpha_{\rm M}^{\rm c}$ and $f_{\rm M}^{\rm c}$ derived from Fig.~\ref{fig:4.3.1} are valid only over SGC34. If we increase the sky fraction to $b \le -30\deg$ covered by the GASS \hi\ survey, we get the same $\alpha_{\rm M}^{\rm c}$ value, but a slightly higher $f_{\rm M}^{\rm c}$ value ($f_{\rm M}^{\rm c}=0.5$). This suggests that the relative strength of \Bt\ with respect to \Bo\ might vary across the sky.

\subsection{Values of model parameters}
\label{choiceparams}

Here we describe how the nine model parameters were found.

We found the mean direction of \Bo\ to be $l_0=73\pdeg5$ and $b_0=23\pdeg5$ from a simple model fit of normalized Stokes parameters for polarization over SGC34 (Sect.~\ref{sec:3.1.2}, Table~\ref{tab:1}). 

As described in Sect.~\ref{sec:cnm_alignment}, using $\NHI^{\rm c}$ as a tracer of the dust polarization angle gives the constraints $\alpha_{\rm M}^{\rm c}=-2.4$ and $f_{\rm M}^{\rm c}=0.4$. The spectral index $\alpha_{\rm M}^{\rm u/w}$ of \Bt\ for the UNM and WNM components is assumed to be same as the $\alpha_{\rm M}^{\rm c}$ value of the CNM component, for simplicity.

Only four model parameters, $\langle \epsilon_{353} \rangle$, $p_0$, $f_{\rm M}^{\rm u/w}$, and $\sigma_{\rm c}$, are adjusted by evaluating the goodness of fit through a $\chi^2$-test, specifically
\begin{equation}
\chi_{XX}^2 = \sum_{\ell_{\rm min}}^{\ell_{\rm max}} \left[ \frac{\dlxx - \mlxx (\langle \epsilon_{353} \rangle, p_0, f_{\rm M}^{\rm u/w}, \sigma_{\rm c})}{\sigma_{\ell}^{XX}} \right]^2
\end{equation}
and
\begin{equation}
\chi^2 = \sum_i^{N_{\rm pix}} \left[ \Dmodel\ - s I_{\rm m, 353} (\langle \epsilon_{353} \rangle, p_0, f_{\rm M}^{\rm u/w}, \sigma_{\rm c}) - o \right]^2 \ ,
\end{equation}
where $XX=\{EE, BB, TE\}$, \dlxx, and $\sigma_{\ell}^{XX}$ are the mean observed dust polarization band powers and standard deviation at 353\,GHz, respectively, \mlxx\ is the model dust polarization power spectrum over SGC34, $N_{\rm pix}$ is the total number of pixels in SGC34, $s$ and $o$ are the slope and the offset of the `T-T' correlation between the \Dmodel\ and $I_{\rm m, 353}$ maps.  We evaluated  $\chi_{XX}^2$ using the six band powers in the range $40 < \ell < 160$.   For $\chi^2$ minimization, we put a flat prior on the mean dust emissivity, $\langle \epsilon_{353} \rangle = (111 \pm 22)\,\mukcmb (10^{20}\,\cm2)^{-1}$ at 353\,GHz, as obtained by correlating \NHI\ with \Dmodel\ map over SGC34. Based on the observed distribution of the polarization fraction (Sect.~\ref{sec:3.1.1}), we put a flat prior on $p_0=(18\pm4)\,\%$.
To match the observed dust amplitude at 353\,GHz, the value of $s$ should be close to 1
and is kept so by adjusting $\langle \epsilon_{353} \rangle$ during the iterative solution of the two equations.

The typical mean values of $\langle \epsilon_{353} \rangle$, $p_0$, $f_{\rm M}^{\rm u/w}$, and $\sigma_{\rm c}$ that we obtained are $121.8\,\mukcmb (10^{20}\,\cm2)^{-1}$, 18.5\,\%, 0.1, and 7.5\kms, respectively, with $\chi^2_{EE}=5.0$, $\chi^2_{BB}=3.3$, and $\chi^2_{TE}=4.9$ for six degrees of freedom (or band-powers).
The precise optimization is not important and is beyond the scope of this paper. 

The model is not particularly sensitive to the value of the final parameter, $\sigma_{\rm u}$, but it is important to have significant column density in each of UNM and WNM so as to have contributions from three layers; we chose $\sigma_{\rm u} = 10 \kms$, for which the average column densities of the phases over SGC34 are rather similar (Sect.~\ref{regionselection}).

Our value of $\langle \epsilon_{353} \rangle$ differs slightly from the mean value $\langle \epsilon_{353} \rangle$ quoted in Table~2 of \citet{planck2013-XVII}, where the \planck\ intensity maps are correlated  with \hi\ LVC emission over many patches within Galactic latitude $b < -25\deg$.  A difference in the mean dust emissivity is not unexpected because our study is focussed on the low column density region ($\NHI\,\le \, 2.7\times10^{20}\,\cm2$) within the \citet{planck2013-XVII} sky region ($\NHI\,\le \, 6\times10^{20}\,\cm2$). 

\begin{figure}
\begin{tabular}{c}
\includegraphics[width=8.8cm]{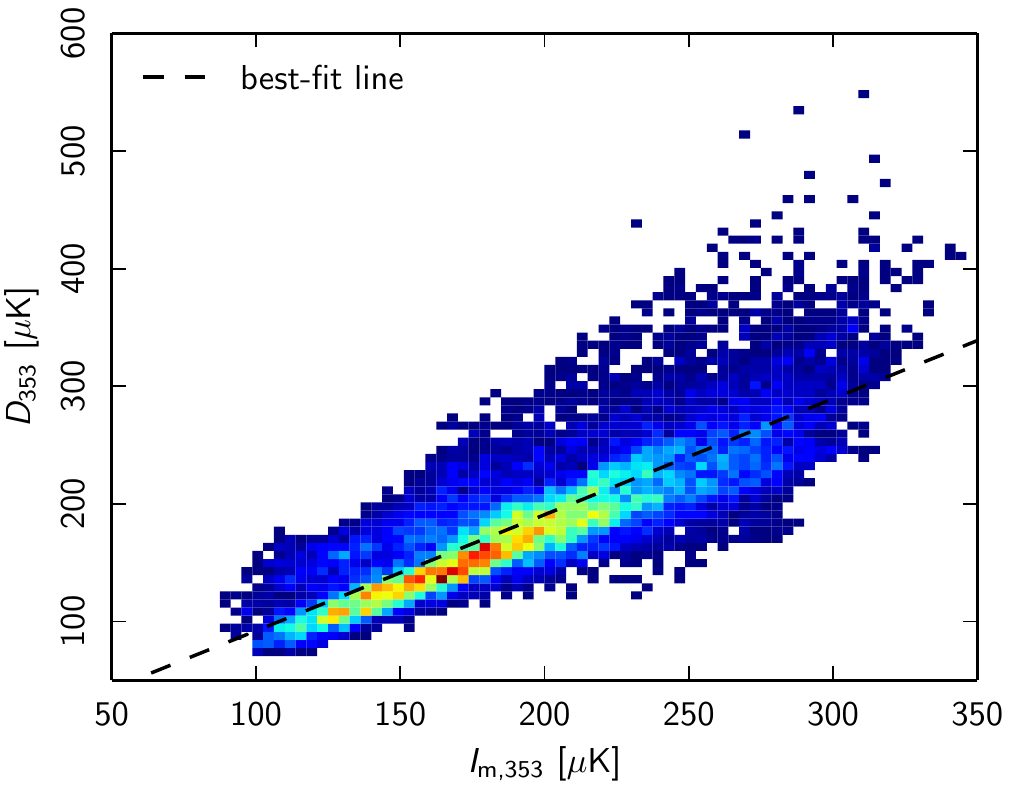}
\end{tabular}
\caption{Correlation of the \planck\ 353\,GHz dust intensity map and the dust model Stokes $I$ map. The dashed line is the best-fit relation.}
\label{fig:dustcorr}
\end{figure}

Fig.~\ref{fig:dustcorr} shows the correlation between the data \Dmodel\ and the model  $I_{\rm m, 353}$ over SGC34. The parameters of the best fit line are $s =1.0$ 
(iteratively by construction)
and $o = -6.6\,\mukcmb$. We compute a residual map $\resdiff \equiv \Dmodel -  I_{\rm m, 353}$ over SGC34. The dispersion of the residual emission with respect to the total \Dmodel\ emission is $\sigma_{\resdiff}/\sigma_{\Dmodel}=0.36$. This means that the residual emission accounts for only 13\,\% ($\sigma_{\resdiff}^2/\sigma_{\Dmodel}^2$) of the total dust power at 353 GHz. We calculate the Pearson correlation coefficients over SGC34 between the residual map and the three \hi\ templates, namely the CNM, UNM, and WNM, finding $-0.09$, $-0.17$, and $0.13$, respectively. The residual might originate in pixel-dependent variations of $\epsilon_{353}$ or possibly from diffuse ionized gas that is not spatially correlated with the total \NHI\ map. \citet{planck2013-XVII} explored these two possibilities (see their Appendix D) and were able to reproduce the observed value of $\sigma_{\resdiff}/\langle \Dmodel \rangle$ at 353\,GHz. 
We do not use the residual emission in our analysis because our main goal is to reproduce the mean dust polarization properties over the SGC34 region. For future work, the pixel-dependent variation of $\epsilon_{353}$, like in the \citet{planck2013-XVII} analysis, could be included so that one could compare the refined dust model with the observed dust polarization data for a specific sky patch (e.g. the BICEP2 field).

\subsection{Monte-Carlo simulations}
\label{sec:monte-carlo}

Using the general framework described in Sect.~\ref{sec:methodology}, we simulate a set of 100 polarized dust sky realizations at 353\,GHz using a set of model parameters, as discussed in Sect.~\ref{sec:4.4}. Because the model of the turbulent GMF is statistical, we can simulate multiple dust sky realizations for a given set of parameters. The simulated dust  intensity and polarization maps are smoothed to an angular resolution of FWHM 1\deg\ and projected on a \healpix\ grid of \Nside=128. These dust simulations are valid only in the low column density regions.

To compare the dust model with the \planck\ observations, we add realistic noise simulations called full focal plane 8 or ``FFP8" \citep{planck2014-a14}. These noise simulations include the realistic noise correlation between pair of detectors for a given bolometer. For a given dust realization, we produce two independent samples of dust Stokes $Q$ and $U$ noisy model maps as 
\begin{align}
Q_{\rm m, 353}^{\rm s_i} (\hat{\vec{n}})  &=  Q_{\rm m, 353}(\hat{\vec{n}})  +  Q_{\rm n}^i  (\hat{\vec{n}})\ ,   \nonumber \\
U_{\rm m, 353}^{\rm s_i} (\hat{\vec{n}})  &=  U_{\rm m, 353} (\hat{\vec{n}}) +  Q_{\rm n}^i (\hat{\vec{n}}) \ ,  \label{eq:5.4}
\end{align}
where $i=1,2$ are two independent samples, and  $Q_{\rm n}^i$ and $U_{\rm n}^i$ are the statistical noise of the subsets of the \planck\ polarization maps (as listed in Table~\ref{tab:1}).

\begin{figure}
\begin{tabular}{c}
\includegraphics[width=8.8cm]{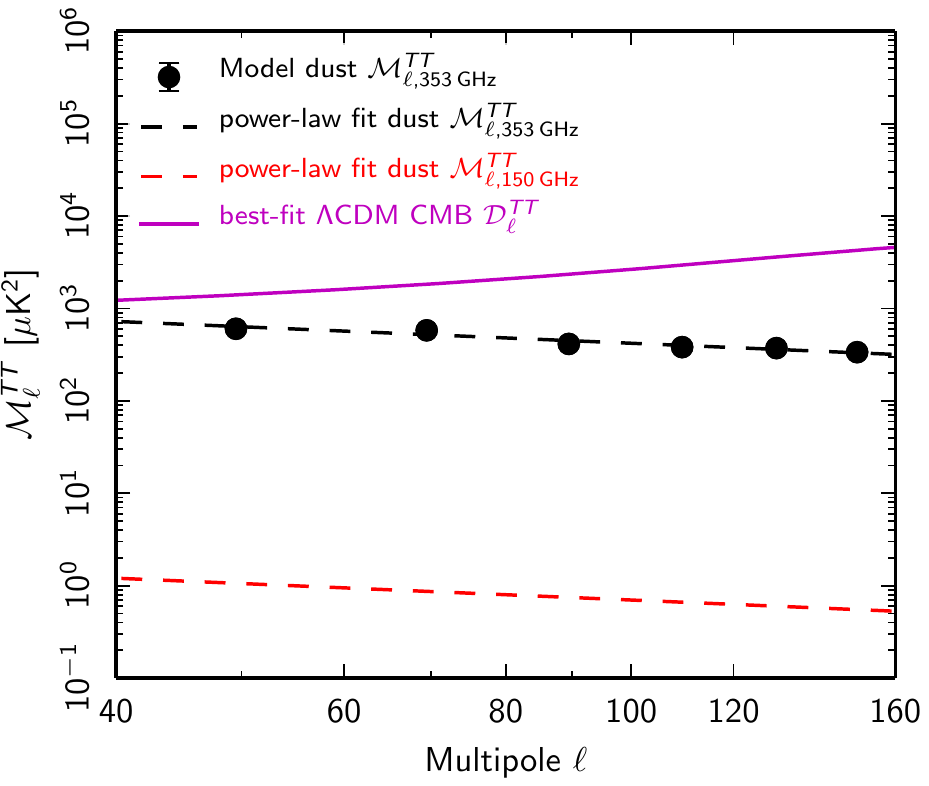}
\end{tabular}
\caption{ Best-fit power law dust model $TT$ power spectra at 353\,GHz (black dashed line) and extrapolated dust power spectra at 150\,GHz (red dashed line) as computed over SGC34. For comparison, we also show the \planck\ 2015 best-fit $\Lambda$CDM expectation curves of the CMB signal. }
\label{fig:TTps}
\end{figure}

\subsection{Comparison of the dust model with \planck\ observations}
\label{sec:results}

In this section we compare the statistical properties of the dust polarization from the dust model with the \planck\ 353\,GHz observations over the selected region SGC34.

We showed in Fig.~\ref{fig:dustcorr} that the model Stokes $I_{353}^{\rm m}$ provides a good fit to the \Dmodel\ map in amplitude.  Here we show that this is the case for the shape of the \dltt\ power spectrum too. We compute the model dust $TT$ power spectrum, $\mltt \equiv \ell (\ell+1) C^{TT}_{\ell, \rm m}/(2\pi)$, over SGC34 in the range $40 < \ell < 600$. The dust \mltt\ power spectrum is well represented with a power-law model, $\mltt \propto (\ell/ 80)^{\alpha_{TT}+2}$, with a best-fit slope $\alpha_{TT}=-2.59\pm 0.02$, as shown in Fig.~\ref{fig:TTps}. Our derived best-fit value of $\alpha_{TT}$ is consistent with that measured in \citet{planck2013-p11} and \citet{planck2014-a15} at high Galactic latitude. For comparison, we also show the $TT$ power spectrum of the CMB signal for the \planck\ 2015 best-fit $\Lambda$CDM model \citep{planck2014-a15}.  The best-fit dust \mltt\ power spectrum extrapolated from 353\,GHz to 150\,GHz, scaling as in Fig.~\ref{fig:3.2} and shown here as a dashed red line, reveals that over SGC34 the amplitude of the dust emission at 150\,GHz is less than 1\,\% compared to the amplitude of the CMB signal. 

Next we show that the dust model is able to reproduce the observed distribution on $p_{\rm d}^2$ and dispersion of $\psi^{\rm R}_{\rm d}$ over SGC34 shown in Figs.~\ref{fig:3.1.1} and \ref{fig:3.1.1a}.  We note that these statistics were not used to optimize the four model parameters in Sect.~\ref{choiceparams}, and so this serves as an important quality check for the model. We compute the square of the polarization fraction, $p_{\rm m}^2$, for the dust model
\begin{equation}
p_{\rm m}^2  =  \left < \frac{Q_{\rm m, 353}^{\rm s_1} \, Q_{\rm m, 353}^{\rm s_2}  \, + U_{\rm m, 353}^{\rm s_1} \, U_{\rm m, 353}^{\rm s_2}}{I_{\rm m, 353}^{2}} \right > \ , 
\end{equation}
where the index `${\rm m}$' again refers to the dust model.  We use the two independent samples to exploit the statistical independence of the noise between them. 
The mean normalized histogram of $p_{\rm m}^2$ and the associated $1\sigma$ error bars are computed over SGC34 using 100 Monte-Carlo simulations.  Returning to Fig.~\ref{fig:3.1.1}, we compare the normalized histogram of $p_{\rm m}^2$ (black circles) with the normalized histogram of $p_{\rm d}^2$ (blue inverted-triangles). The \planck\ instrumental noise (FFP8 noise) added in the dust model accounts nicely for the observed negative values of $p_{\rm m}^2$ and also contributes the extension of the $p_{\rm m}^2$ distribution beyond the input value of $p_0^2$. The value of $p_0$ found in this paper is close to the value of 19\% deduced at 1\deg\ resolution over intermediate and low Galactic latitudes \citep{planck2014-XIX}.

We also compute the dispersion of the polarization angle $\psi_{\rm m}^{\rm R}$ for the dust model. We follow the same procedure as discussed in Sect.~\ref{sec:3.1.3} and compute the angle
\begin{equation}
\psi_{\rm m}^{\rm R} = \frac{1}{2} \tan^{-1} ( - U_{\rm m, 353}^{\rm R}, Q_{\rm m, 353}^{\rm  R} ) \ ,   
\end{equation}
where $ Q_{\rm m, 353}^{\rm R}$ and $U_{\rm m, 353}^{\rm R}$ are rotated Stokes parameters with respect to the local direction of the large-scale GMF. We make the normalized histogram of $\psi_{\rm m}^{\rm R}$ over SGC34. The mean normalized histogram and associated $1\sigma$ error bars of $\psi_{\rm m}^{\rm R}$ for the dust model are computed from 100 Monte-Carlo simulations.  Returning to Fig.~\ref{fig:3.1.1a}, we compare the dispersion of $\psi_{\rm m}^{\rm R}$ (black points) with the dispersion of  $\psi^{\rm R}_{\rm d}$ (blue points). The $1\sigma$ dispersion of $\psi_{\rm m}^{\rm R}$ derived from the mean of 100 dust model realizations is $21\pdeg0\pm0\pdeg7$, slightly larger than for the \planck\ data ($15\pdeg0\pm0\pdeg4$, Sect.~\ref{sec:3.1.2}). 
This could come from the simple modelling assumptions used in our analysis. In particular, the histograms of $p_{\rm d}^2$ and $\psi_{\rm d}^{\rm R}$ depend on the dust modelling at low multipoles ($\ell < 40$), which are not constrained by our data fitting. Varying some of the model parameters such as $\alpha_{\rm M}^{\rm u/w}$, different $R$ for different ISM phases, or introducing a low-$\ell$ cutoff in the spectral index of the \Bt\ \citep{Cho:2002, Cho:2010} might provide a better fit of these histograms 
to the \planck\ data. Due to the limited sky coverage over SGC34 and residual systematics in the publicly available Planck PR2 data at low multipoles  \citep{planck2014-a10}, we are unable to test some of these possibilities within our modelling framework.

Finally, we compute power spectra of the dust model over SGC34 using the cross-spectra of two independent samples as described in Sects.~\ref{sec:3.1.3} and \ref{sec:monte-carlo}.  Mean dust polarization power spectra of the noisy dust model, \mlee, \mlbb, and \mlte, over the multipole range $40 < \ell < 160$ are calculated from 100 Monte-Carlo simulations and presented in the right column of Fig.~\ref{fig:3.2}. The associated $1\sigma$ error bar per $\ell$ bin is also derived from the Monte-Carlo simulations.  All of the dust polarization power spectra are well represented by a power law in $\ell$.  We fit the model spectra with a power-law model over the multipole range $40 < \ell < 160$. The best-fit values of the exponents and then of the amplitudes with exponent $-2.3$ are quoted in Table~\ref{table2}. Because of the $\chi^2$ optimization of the four model parameters (Sect.~\ref{choiceparams}), these power spectra agree well with the observed power spectra over SGC34.

\begin{figure*}
\includegraphics[width=17.8cm]{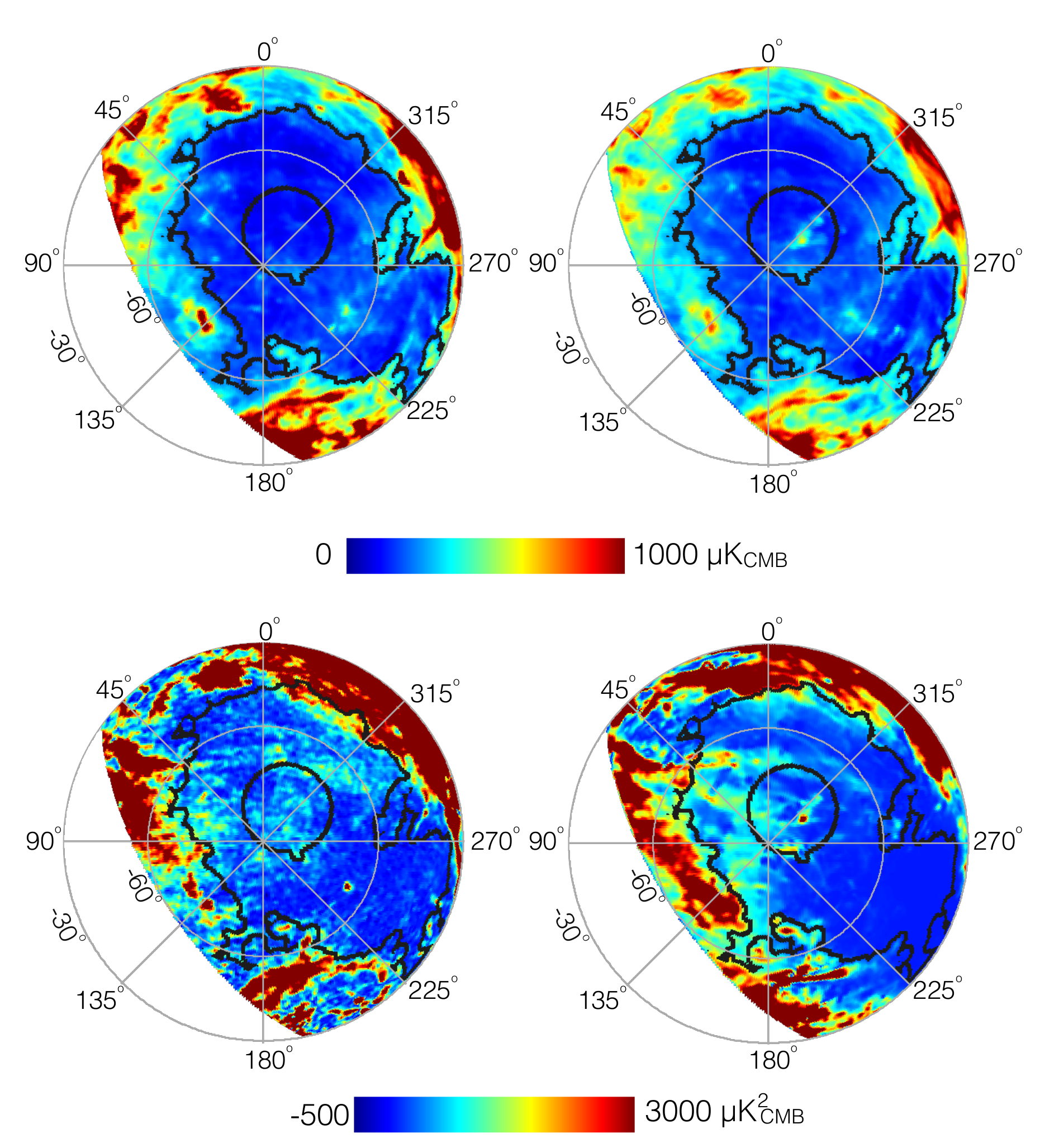}
\caption{\textit{Left column}: Orthographic projections of the \planck\ 353\,GHz \Dmodel\  (Stokes $I$) map (\textit{top row}) 
and the square of the polarization intensity $P^2$ (\textit{bottom row}), at 1\deg\ resolution over the southern Galactic cap covered by the GASS \hi\ survey, with the selected region SGC34 outlined by a black contour. The same coordinate system is used as in Fig.~\ref{fig:4.2.1}. \textit{Right column}: Similar to left panels, but for a realization of the 353\,GHz dust model. Judged visually, the dust model is able to reproduce roughly the observed dust polarization sky over SGC34, and indeed over the entire field shown.}
\label{fig:5.2}
\end{figure*}

\section{Predictions from the dust model}
\label{sec:prediction}

The simulated dust emission maps with FFP8 noise approximate the \planck\ 353\,GHz data well over SGC34. In this section we use \emph{noiseless} dust simulations to make a few predictions concerning the foreground dust $B$-mode that are important for the search for the CMB $B$-mode signal. These predictions include the $TB$ and $EB$ correlation for the dust emission and the statistical variance of the dust $B$-modes at the power spectrum level.

But first we compare the spatial distribution of Stokes $I$ and the polarization intensity ($P^2$) of the noiseless dust model with the \planck\ 353\,GHz data over the southern Galactic cap. The comparison between one realization of the dust model and the \planck\ data is presented in Fig.~\ref{fig:5.2}. The dust simulation is derived completely from the GASS \hi\ data and a phenomenological description of the large-scale and turbulent components of the GMF with parameters optimized over SGC34, a subregion of the SGC.  We find a good match between the data and the dust model over SGC34, outlined by the black contour. Outside of SGC34,  the brightest emission features appear stronger in the \planck\ 353\,GHz data because the \hi-based dust model does not account for dust emission associated with $\mathsc{H}_2$ gas and with \hi\ gas that is too cold to result in significant net emission.

\subsection{Dust $TB$ and $EB$ correlation}
\label{sec:tbandeb}

The $TB$ and $EB$ cross-spectra vanish for the CMB signal in the standard $\Lambda$CDM model \citep{Zaldarriaga:1997}. Here we test whether the corresponding dust model $TB$ and $EB$ cross-spectra vanish or not in the low column density region SGC34. To do this we simulate 100 Monte-Carlo realizations of \emph{noiseless} dust skies and compute the cross-spectra for each realization. 

\begin{figure}
\begin{tabular}{cc}
\includegraphics[width=8.8cm]{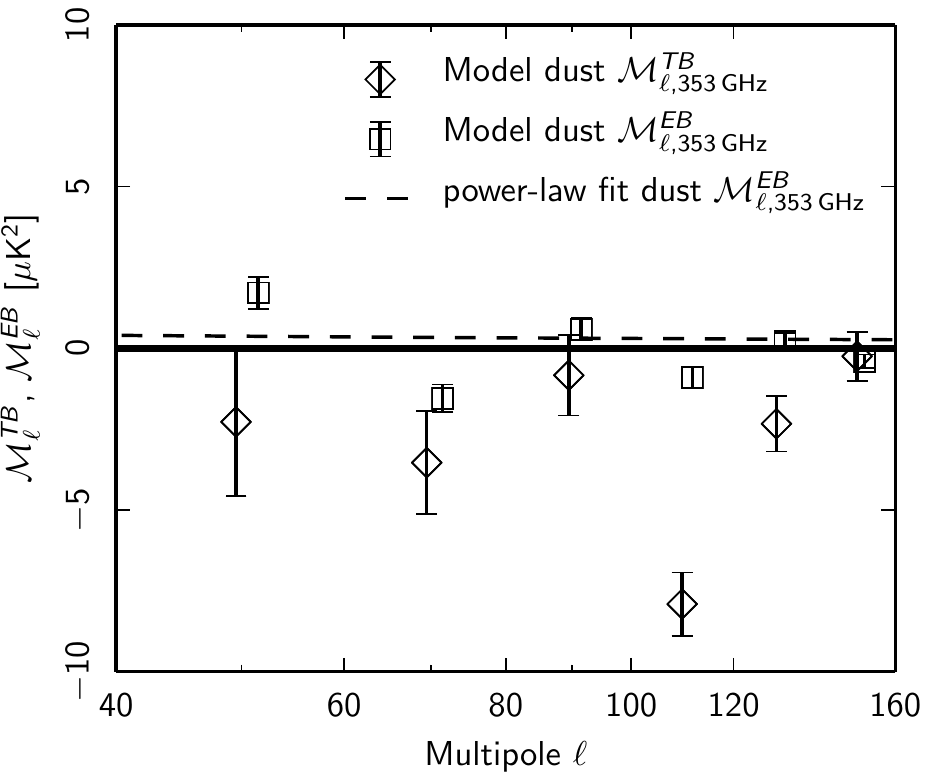}\\
\end{tabular} 
\caption{Predicted dust model power spectra \mltb\ (diamonds) and \mleb\  (squares) over SGC34 computed using 100 noiseless Monte-Carlo dust realizations.  Dashed curve gives a power-law fit to \mleb\ (see Sect.~\ref{sec:tbandeb}).}
\label{fig:7.1}
\end{figure}

The mean model dust \mleb\ spectrum over the multipole range $40 < \ell < 160$ is presented in Fig.~\ref{fig:7.1}.  Similar to \citet{Abitbol:2016}, we fit the $EB$ power spectrum with a power-law model, $\mleb = A_{EB}(\ell/80)^{\alpha_{EB}+2}$ with a fixed slope $\alpha_{EB}=-2.3$ and extrapolate it to 150\,GHz using the scaling factor $s_{150/353}$. Our dust model predicts $A_{EB}=0.31\pm 0.13\, \mukcmbsq$ ($2.4\,\sigma$ level) at 353\,GHz. The best-fit value of $A_{EB}$ at $\ell=80$ translates into an $EB$ amplitude of $0.52\times10^{-3} \mukcmbsq$ at 150\,GHz, consistent with the results presented in Table~A1 of \citet{Abitbol:2016} over the BICEP2 patch (1.3\,\% of the cleanest sky region). Because our dust simulations are noiseless, we can determine the amplitude of the $EB$ spectrum with relatively small error bars. This non-zero dust $EB$ spectrum towards the SGC34 region can produce a spurious $B$-mode signal, if not taken into account in the self-calibrated telescope polarization angle \citep{Abitbol:2016}. Comparing our result with \citet{Abitbol:2016} indicates that the amplitude of spurious $B$-mode signal is negligible at $\nu~\le~150$\,GHz, but becomes important at dust frequency channels $\nu~>~217$\,GHz.

On the other hand, the mean model dust \mltb\ spectrum is mostly negative, but consistent with zero within the $3\sigma$ error bar for each multipole bin except for $\ell=110$. The amplitude  ${\cal M}_{\ell=110,353\,\text{\rm GHz}}^{TB}=-7.9\,\mukcmbsq$ (or ${\cal M}_{\ell=110,150\,\text{\rm GHz}}^{TB}=-0.013\, \mukcmbsq$ is consistent with the \citet{Abitbol:2016} value over the BICEP2 patch (1.3\,\% of the cleanest sky region). The origin of the negative \mltb\ at $\ell=110$ is currently unknown. However, the significance of the negative \mltb\ amplitude goes down with the choice of binning $\Delta {\ell} > 30$. It would be interesting to check our predicted model dust \mleb\ and \mltb\ spectra with the upcoming BICEP2 dust power spectra at 220 GHz.
%

\begin{figure}
\begin{tabular}{cc}
\includegraphics[width=8.8cm]{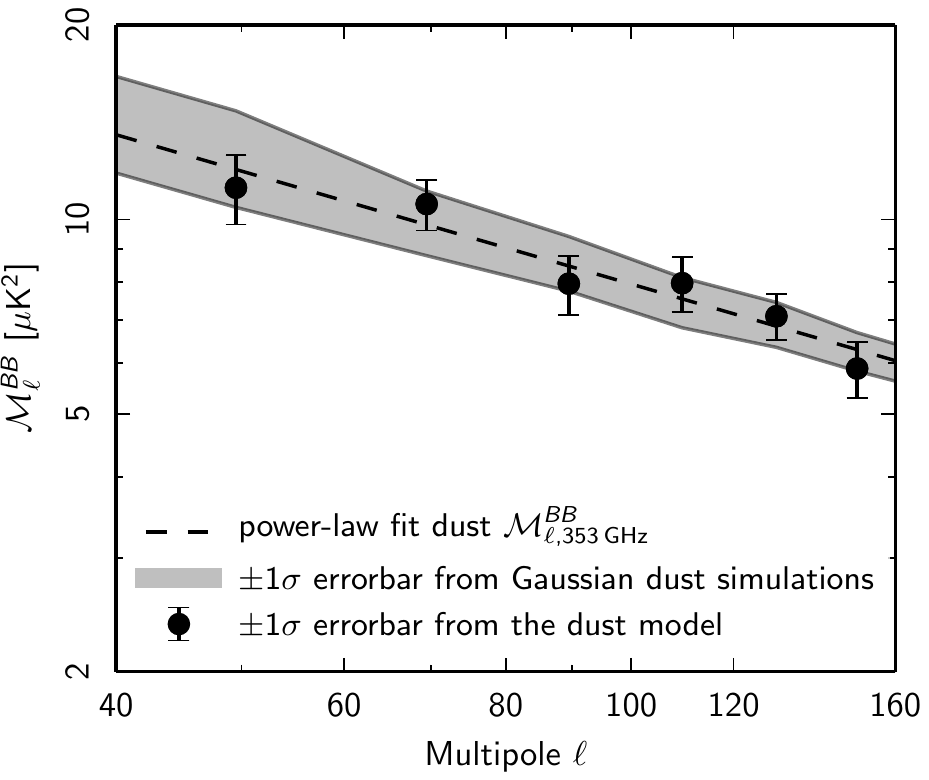}
\end{tabular} 
\caption{Comparison of the cosmic variance of the dust model \dlbb power spectrum with the one derived from pure Gaussian dust simulations. The statistical variance at low multipoles ($\ell < 80$) results from the fixed mean direction of the large-scale GMF (see Sect.~\ref{sec:statistical_variance}). }
\label{fig:7.1a}
\end{figure}

\subsection{Statistical variance of the dust model}\label{sec:statistical_variance}

In the analysis in \citet{pb2015}, Gaussian random realizations of the dust sky are simulated with a spatial power law ${\cal M}_{\ell} \propto \ell^{-0.42}$ at 353\,GHz (and scaled to other microwave frequencies using a MBB spectrum with $\beta_{\rm d}=1.59$ and $T_{\rm d}=19.6$\,K). In contrast to that Gaussian dust model, our dust model produces non-Gaussianity because it is based on the GASS \hi\ data. Here we test to what degree the statistical variance of the new dust simulations is compatible with the pure Gaussian model approximation.

For this test we proceed as in Sect.~\ref{sec:results} by estimating the mean dust \mlbb\ power spectrum at 353\,GHz over the multipole range $40 < \ell < 160$ and the $1\sigma$ standard deviations from 100, in this case noiseless, dust sky realizations over SGC34. These dust model results are shown in Fig.~\ref{fig:7.1a} as black points, with a power-law fit as the dashed line. Then we assume the hypothesis of Gaussian and statistical isotropy and simulate 100 Gaussian dust sky realizations. The $1\sigma$ error range for the BB power from these Gaussian dust realizations is shown as the shaded region in Fig.~\ref{fig:7.1a}.

Comparing these results, our dust simulation band powers shown by the black symbols have less statistical variance at low multipoles $\ell < 80$ compared to the variance from the Gaussian sky simulations. This reflects the fact that the model is not constrained exclusively by the dust polarization power spectra.  In our dust model, the mean GMF orientation is fixed and the polarization angle of the CNM layer is determined by the GASS \hi\ data. However, at high multipoles $\ell > 80$, the effect of \Bt\ starts dominating over \Bo, leading to a statistical variance that is comparable to or higher than that from the Gaussian sky simulations.  

In summary, our dust model predicts smaller statistical variance at multipoles $\ell < 80$ compared to the Gaussian dust approximation. Such an effect in the variance is not seen in the \citet{Vansyngel:2016} dust model over the same multipole range for the LR regions defined in \citet{planck2014-XXX}. This is because the amplitudes of the dust polarization power in the \citet{Vansyngel:2016} analysis are dominated by the brightest sky areas within these larger regions, whereas by its selection criteria the SGC34 region is more homogeneous.

\section{Astrophysical interpretation}\label{sec:interpretation}

In this section we present an astrophysical interpretation of our dust modelling results.

In our analysis the alignment of the CNM structures with \Bpos\ sets the spectral index of the \Bt\ power spectrum, $\alpha_{\rm M}^{\rm c}=-2.4$. The \citet{Vansyngel:2016} parametric dust model reports $\alpha_{\rm M}=-2.5$ for each polarization layer using the CIB-corrected 353 GHz GNILC dust map \citep{planck2016-XLVIII}. These two complementary analyses show that the spectral index of the \Bt\ power spectrum is very close to the slope of the polarized dust power spectra (\dlee, \dlbb, and \dlte).

\begin{figure}
\begin{tabular}{cc}
\includegraphics[width=8.8cm]{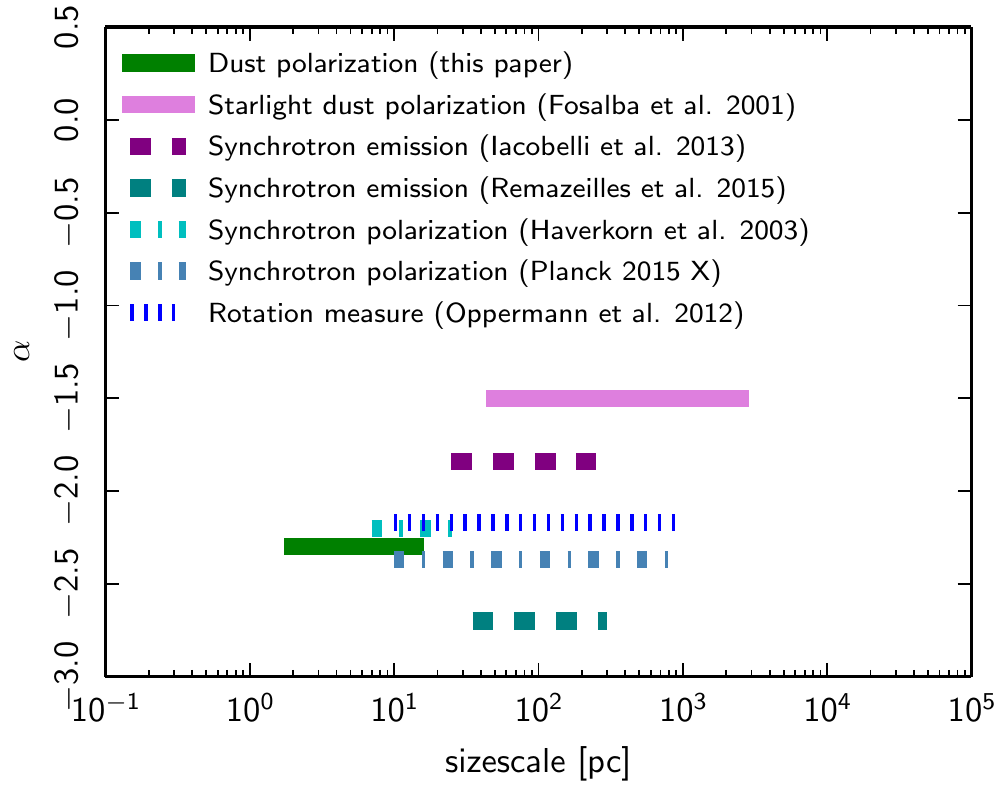}
\end{tabular} 
\caption{Summary of the observed spectral indices of the power spectrum as a function of length scales (in pc), from dust polarization (solid), synchrotron emission (dashed), synchrotron polarization (dashed-dot), and rotation measures (dotted). See Sect.~\ref{sec:interpretation} for a detailed description. }
\label{fig:8.1}
\end{figure}

Our value of the spectral index $\alpha_{\rm M}$ is significantly higher (the spectrum is less steep) than the value of $-11/3$ from the Kolmogorov spectrum. In Fig.~\ref{fig:8.1}, we compare our result with the spectral indices obtained in earlier studies using starlight polarization \citep{Fosalba:2002}, synchrotron emission \citep{Iacobelli:2013, Remazeilles:2015}, synchrotron polarization \citep{Haverkorn:2003,planck2014-a12}, and rotation measures \citep{Oppermann:2012}. For each data set we convert the $\ell$ range to physical sizes.  We used a scale-height of 1\,kpc for the warm ionized medium and the synchrotron emission. For the study of SGC34 here, the maximum distance of the dust emitting layer is $L_{\rm max}=100-200$\,pc, a range between estimates of the typical distance to the local bubble \citep{Lallement:2014} and of the dust emission scale height \citep{Drimmel:2001}. All of the values are larger than the spectral index of the Kolmogorov spectrum. In several papers this is explained by introducing an outer scale of turbulence close to the measured physical sizes. However, the current data do not show a consistent systematic trend of spectra steepening towards smaller sizes.
Such a trend might be difficult to see combining different tracers of the magnetic field. For dust polarization, we fit the \planck\ observations over the multipole range $40 < \ell < 160$, which corresponds to physical scales of $2-15$\,pc on the sky. At larger scales, the spectral index from starlight dust polarization (interstellar polarization) is flatter.  If this interpretation is correct, then we expect to see further flattening of the \Bt\ power spectrum at low multipoles (see Figure~20 of \citealt{planck2014-a10}). Such flattening, which we neglected in this analysis, could possibly account for the slight differences in the histograms of $p^2$ and $\psi^{\rm R}$ between the \planck\ data and the dust model (Figs.~\ref{fig:3.1.1} and ~\ref{fig:3.1.1a}).

With our modelling assumption of the same dust alignment parameter ($R$; see Sect.~\ref{sec:4.1.1}) for each \hi\ component, we find that the value of $f_{\rm M}^{\rm c}$ is significantly larger than the value of $f_{\rm M}^{\rm u/w}$. The size of $f_{\rm M}^{\rm c}$ follows from the statistics of the \hi\ CNM map because we assume that \Bpos\ is aligned with CNM structures (Sect.~\ref{sec:cnm_alignment}). From  $f_{\rm M}^{\rm c}=0.4$, the value of $f_{\rm M}^{\rm u/w}$ (0.1) is set by the amplitudes of the observed polarization power spectra (Sect.~\ref{choiceparams}). Through a consistency check we find that the dispersion of $\psi_{\rm m}^{\rm R}$ depends on the $f_{\rm M}^{\rm u/w}$ value. If we increase the size of $f_{\rm M}^{\rm u/w}$, the dispersion of  $\psi_{\rm m}^{\rm R}$ becomes larger than the observed dispersion of $\psi_{\rm d}^{\rm R}$.  Thus our modelling indicates that the CNM component is more turbulent than the UNM/WNM components. This is the first time that this behaviour has been observed in the \planck\ polarization data. \citet{Kalberla:2016} and \citet{Heiles:2003} have shown that the CNM has the largest sonic Mach number. If both the sound and Alfv\'{e}n speed change by a same factor ($\simeq10$) between the CNM and WNM, we also expect a higher Alfv\'{e}nic Mach number for the CNM than the WNM component \citep{Hennebelle:2006}. This would fit with the larger $f_{\rm M}$ value in the CNM.

Alternatively, the lower value of $f_{\rm M}$ for the WNM component could result from our modelling that reduces the LOS integration to a small number of discrete layers. \citet{planck2016-XLIV} argued that the number of layers ($N_{\rm layers}$) is related to the multiphase structure of the ISM and to the correlation length of the GMF. For the CNM, dust polarization traces the GMF at discrete locations along the LOS. For the WNM, dust polarization samples the variations of the GMF orientation along the LOS. For that ISM phase,  $N_{\rm layers}$ is coming from the correlation length of the GMF that depends on the spectral index of the \Bt\ power spectrum. For a typical value of $\alpha_{\rm M}=-2.4$,  the value of $N_{\rm layers}$ is 6 \citep{planck2016-XLIV}. If that reasoning applies, our value of $f_{\rm M}^{\rm w}$ is effectively a factor $\sqrt{N_{\rm layers}}$  smaller than the true value. The same argument might hold for the UNM component too.

Various alignment mechanisms, such as paramagnetic, mechanical, and radiative torques, have been proposed to explain the efficiency of the alignment of dust grains with respect to the magnetic field (see \citealt{Lazarian:2003, Andersson:2015} for a review). \citet{Draine:1996, Draine:1997} argued that the radiative torques mechanism is the dominant process in the diffuse ISM. We have assumed that the degree of grain alignment and the dust polarization properties are the same in all  ISM phases, but they might differ.  We investigated this possibility explicitly in our model by keeping fixed $p_{\rm dust}$ and $f_{\rm M}^{\rm u/w}=0.4$ and varying the alignment parameter in the cold phase ($R_{\rm c}$) and warm phase ($R_{\rm w}$). If we lower $R_{\rm w}$ compared to $R_{\rm c}$, the dispersion of $\psi_{\rm m}^{\rm R}$ becomes broader than the best-fit model presented in this paper. While a lower value $R_{\rm c}$ compared to $R_{\rm w}$ provides a better fit to the dispersion of $\psi_{\rm d}^{\rm R}$, such a model does not reproduce the $E-B$ asymmetry in the power spectrum amplitudes. In summary, within our modelling framework with three \hi\ layers, changes in the (relative) degree of alignment cannot account for the lower value of $f_{\rm M}^{\rm u/w}$.

Our values of $f_{\rm M}^{\rm c}$ and $f_{\rm M}^{\rm u/w}$ are lower than the one reported by \citet{planck2016-XLIV} and \citet{Vansyngel:2016}. This difference follows from the model assumptions. We account for the multiphase structure of the ISM and find $f_{\rm M}^{\rm c} > f_{\rm M}^{\rm u/w}$, while \citet{planck2016-XLIV} and \citet{Vansyngel:2016} keep the same dust total intensity  map and  $f_{\rm M}$  value for all of the polarization layers. As discussed in \citet{planck2016-XLIV}, the best fit values of $f_{\rm M} $  and $p_0$ increase with $N_{\rm layers}$. In our model, the turbulence is most important in the CNM layer and our  value of  $f_{\rm M}^{\rm c} = 0.4$ is in agreement with the best fit value  in \citet{planck2016-XLIV} for their one layer model. We also note that  our  $f_{\rm M}^{\rm c}$ value applies to the SGC34 region, which by definition is a low column density region ($\NHI\,\le \, 2.7\times10^{20}\,\cm2$). Our data analysis suggests that the $f_{\rm M}^{\rm c}$ value might be higher for the whole SGC region. In our model, we put the mean large-scale GMF in both the UNM and WNM components, thus affecting both $\psi^{\rm u}$ and $\psi^{\rm w}$. This feature of the model can be investigated with numerical simulations. Using a simulation of supernova-driven turbulence in the multiphase ISM, \citet{Evirgen:2016} suggest that the mean field might reside preferentially in the WNM.

\section{Discussion and summary}
\label{sec:concl}

We have constructed a phenomenological dust model based on the publicly available GASS \hi\ data and an astrophysically motivated description of the large-scale and turbulent Galactic magnetic field.  The two main ingredients of the model are: (1) fluctuations in the GMF orientation along the LOS and (2) a perfect alignment of the CNM structures and \Bpos. We model the resulting LOS depolarization by replacing the integration along the LOS by the sum of three distinct components that represent distinct phases of the diffuse neutral atomic ISM. By adjusting a set of nine model parameters suitably, we are able to reproduce the observed statistical properties of the dust polarization in a selected region of low column density ($\NHI\,\le \, 2.7\times10^{20}\,\cm2$) that comprises 34\,\% of the southern Galactic cap (SGC34). This dust model is valid only over SGC34.

Unlike PSM dust templates, this dust model is not noise-limited in low column density regions and can be used to predict the average dust polarization at any sky position over the selected region SGC34 at any frequency. This new dust model should be useful for simulating realistic polarized dust skies, testing the accuracy of component separation methods, and studying non-Gaussianity.

Recently,  \cite{Tassis:2015} and \citet{Poh:2016} showed that multiple dust components along the LOS with different temperatures (in general SEDs) and polarization angles lead to frequency decorrelation in both amplitude and polarization direction. Our dust modelling framework allows us to introduce frequency decorrelation of the dust polarization readily by adopting different dust SEDs for the three \hi\ components and then extrapolating the dust model to other microwave frequencies. With the current \planck\ sensitivity the current constraint on the dust frequency decorrelation is only marginal at high Galactic latitude \citep{planck2016-L}.  Although we could use the dust model to explore the frequency decorrelation of the polarized dust emission, this is beyond the scope of this paper.

The main results of the paper are summarized below. 

\begin{itemize}

\item The observed distributions of $p^2$ and $\psi$ over SGC34 can be accounted for by a model with three ISM components: CNM, WNM, and UNM. Each ISM phase has different density structures and different orientations of the turbulent component of the GMF. 

\item  The slopes of \dlee, \dlbb, and \dlte\ angular power spectra over the multipole range $40 < \ell < 160$ are represented well with the same power-law exponent, which is close to the spectral index of the turbulent magnetic field, $\alpha_{\rm M}=-2.4$. 

\item The positive \dlte\ correlation and $E$-$B$ power asymmetry over SGC34 can be accounted for by the alignment of the CNM structures with \Bpos. 

\item The $E$-$B$ power asymmetry for small sky patches depends on the fraction of the total dust emission in the CNM, UNM, and WNM phases. We expect to find the ratio $\dlbb/\dlee$ close to 1 where the UNM/WNM structures dominate over the CNM structures.

\end{itemize}

\begin{acknowledgements}

The research leading to these results has received funding from the European Research Council grant MISTIC (ERC-267934). Part of the research described in this paper was carried out at the Jet Propulsion Laboratory, California Institute of Technology, under a contract with the National Aeronautics and Space Administration. U.H. acknowledges the support by the Estonian Research Council grant IUT26-2, and by the European Regional Development Fund (TK133). The Parkes Radio Telescope is part of the Australia Telescope, which is funded by the Commonwealth of Australia for operation as a National Facility managed by CSIRO. 

Some of the results in this paper have been derived using the \healpix\ package. Finally, we acknowledge the use of \planck\ data available from \Planck\ Legacy Archive (\url{http://www.cosmos.esa.int/web/planck/pla}).

\end{acknowledgements}

\bibliographystyle{aa}


\end{document}